\documentclass[fleqn,10pt]{wlscirep}
\usepackage[utf8]{inputenc}
\usepackage[T1]{fontenc}
\usepackage{float}

\title{FAIR EVA: Bringing institutional multidisciplinary repositories into the FAIR picture}

\author[1,$\dag$]{Fernando Aguilar Gómez}
\author[2,$\dag$]{Isabel Bernal}
\affil[1]{CSIC (Spanish National Research Council), IFCA, Santander, Spain}
\affil[2]{CSIC (Spanish National Research Council), DIGITAL.CSIC, URICI, Madrid, Spain}

\begin{abstract}
The FAIR Principles are a set of good practices to improve the reproducibility and quality of data in an Open Science context. Different sets of indicators have been proposed to evaluate the FAIRness of digital objects, including datasets that are usually stored in repositories or data portals. However, indicators like those proposed by the Research Data Alliance are provided from a high-level perspective that can be interpreted and they are not always realistic to particular environments like multidisciplinary repositories. This paper describes FAIR EVA, a new tool developed within the European Open Science Cloud context that is oriented to particular data management systems like open repositories, which can be customized to a specific case in a scalable and automatic environment. It aims to be adaptive enough to work for different environments, repository software and disciplines, taking into account the flexibility of the FAIR Principles. As an example, we present DIGITAL.CSIC repository as the first target of the tool, gathering the particular needs of a multidisciplinary institution as well as its institutional repository.

\end{abstract}
\begin{document}

\flushbottom
\maketitle

\thispagestyle{empty}

\section*{Introduction}
The four FAIR Principles \cite{Wilkinson2016} (Findable, Accessible, Interoperable and Reusable) are oriented to promote Open Science practices, stimulating collaboration among scientists from different disciplines. Many institutions, research performing and research funding organizations are trying to adopt the principles to make their scientific production transparent, find new ways to develop knowledge and accelerate the making of global and interconnected information infrastructure. Far from being a mere fashionable trend, intergovernmental institutions and research funders like the European Commission are including specific requirements related to the FAIR principles in their agendas. For instance, since Horizon2020 Programme, a growing number of projects have been requested to prepare a complete Data Management Plan based on the EC template, which requires specific details on how the data will be FAIR. Furthermore, within the context of the European Open Science Cloud (EOSC), a specific “FAIR Working Group” aimed at providing recommendations on the implementation of Open and FAIR practices for researchers in Europe \cite{eoscsecretariat_fair}. This group has released different documents including recommendations on how to adopt and implement the FAIR principles in research data and research objects in general. The activities from this working group are being extended within the EOSC Association thanks to the FAIR Metrics and Data Quality Task Force \cite{fair_taskforce}. The report Turning FAIR into a Reality \cite{doi/10.2777/54599} already mentioned the need for metrics and tools to measure FAIRness while another significant recommendation called for avoiding reinventing the wheel since there have been different initiatives working on extending the four FAIR Principles to be applied in real research world. For instance, FAIRmetrics \cite{Wilkinson2018} emerged in close connection with one of the earlier tools, the FAIR Evaluation Services, and other broadly known FAIR data metrics and related criteria have been defined within the context of the Research Data Alliance (RDA), in particular, the FAIR Data Maturity Model Working Group. In 2019 this group released a set of FAIR criteria \cite{rda2020fair} to be fulfilled by research objects like datasets, and it should be the starting point to define any new FAIR metrics as they develop a common set of core assessment criteria for FAIRness. However, this field is evolving continuously and some revisions will need to be done frequently.

Based on the FAIR indicators, there are already evaluation methods and tools providing an estimation on how FAIR a given digital object is \cite{fair_assist}. The Science enterprise is complex and different disciplines have different ways to implement the FAIR Principles and related indicators. Therefore, most of those tools and underlying metrics have focused on solving the problems from a specific domain, or they are too generalist to provide a concrete set of results that allows understanding the actual FAIRness level. There have been several attempts to analyze differences and similarities among such tools and metrics \cite{bahim_christophe_2019_3629618} \cite{Knaisl}.

In this dynamic landscape, one of the most problematic -and to a great extent neglected- environments to apply these tools is the one populated with institutional and multidisciplinary repositories. Large national research-intensive institutions like the Spanish National Research Council (CSIC from now on) group multiple research centres from many different disciplines, which adopt best practices in terms of data management. However, this can be a problem since not every single discipline understands the same about FAIR Principles and holds different views as to the best way to manage research data. At the same time, institutional multidisciplinary repositories have a long life record and were born in the early days of the Open Access movement. In general, they have diversified their agendas and services over time as Open Science has taken root globally. In addition, institutional repositories are bound by several international and national guidelines and good practices that aim to foster harmonization, standardization and interoperability. Amongst well-known guidelines in this respect, it is worth mentioning COAR Community Framework for Good Practices in Repositories \cite{coarweb}, OpenAire Guidelines for Repositories \cite{openaire_guide} and BASE Golden Rules for Repository Managers \cite{base_guide}, to name but a few. This paper presents FAIR EVA (Evaluator, Validator \& Advisor), a developed solution that takes into account the actual way of working of institutional repositories so that assessing compliance with FAIR Principles results in a useful exercise. This is a gap that EOSC FAIR Working Groups are already aware of and suggest taking into account \cite{doi/10.2777/986252}. This paper describes the first target of FAIR EVA tool, namely, that of focusing on multidisciplinary repositories, without neglecting opportunities to adapt it to other scenarios.

%Esto igual se puede quitar
Another potential problem is the continuous and exponential growth of not only data but also other research outputs and digital objects like software which are also worthy of being included in such assessments. In addition, as a research project evolves new versions of a given output may get published and so do associated metadata and supplementary materials, meaning that their FAIRness degree may change, too. Although there are some manual services to measure the FAIRness level \cite{Clarke2019} \cite{Bonello2022}, their use is discouraged since they can not address the evaluation of many digital objects quickly. On the contrary, automatic tools should be capable of providing a set of measurements or results based on the provided indicators, showing how FAIR principles can be better adopted. The aforementioned EOSC FAIR Working Group suggests not recommending only one tool, but trying to compare in order to identify biases and take into account specific contexts. For example, the FAIR evaluation services \cite{Wilkinson2019} execute a set of metrics to give a result on how FAIR a given dataset is or else, which criteria are correctly fulfilled and which are not and why, thus providing guidelines for improvement.

The final goal of adopting the FAIR Principles and producing FAIR data and other research products is to enhance their re-usability, stimulate interdisciplinary research and promote Open Science by providing more transparency on the way science is conducted. Open Science has proved as the way to make Science more efficiently and faster to yield results. Although it could be improved \cite{Homolak2020}, the pandemic of COVID-19 has shown that Open Science practices can incentivize and facilitate the collaboration and the finding of solutions for given problems. Data portals and computing infrastructures of information sharing have been appearing to help scientist face this global problem. The terms Open Science and FAIR data are still pretty new for quite many researchers in the global community, and awareness-raising, capacity building and training are still very much needed to enable a systemic change. FAIRness assessment tools that provide feedback and tips on how specific metrics or indicators can be improved can help researchers be more aware of the importance of data management and how to adopt FAIR Principles to their research field. Furthermore, FAIRness assessment tools can be instrumental for repository administrators and repository software developers to identify technological gaps. The tool presented here combines these expected features of being automatic, scalable, educational and supportive in order to enable deeper compliance of FAIR Principles in the reality of scientific institutional and disciplinary repositories.

\section*{Goals}
This paper aims to present the main characteristics of FAIR EVA, a new service-oriented to provide feedback for researchers, repositories administrators and developers in order to let them know the current level of FAIRness of digital objects. The goal is to provide a tool flexible enough to be adapted to different types of repositories and researcher profiles, as well as make available mechanisms for specific systems like institutional repositories to evaluate the FAIRness of their digital objects within their own institutional context.

In particular, this tool has been developed taking into account the context and reality of CSIC \cite{csicweb}, the largest public research performing organization in the country and the 5th research organization with more European Commission funded H2020 actions, comprised of staff over 12,000 people, 120 research centres and institutes across 3 broad research areas (Life, Matter and Society), about 1,500 research groups and 35 institutional interdisciplinary thematic platforms, as well as the context of its institutional repository DIGITAL.CSIC which organizes, describes and opens access to institutional research outputs. Starting from this well-known data service will facilitate the adaptation of FAIR EVA to other production environments.

\subsection*{Technical implementation of FAIR indicators}
Currently, the best-agreed set of FAIR indicators has been produced by the RDA FAIR Data Maturity Model WG \cite{rda2020fair}, composed of several hundreds researchers and professionals from diverse disciplines. Furthermore, the creation of these indicators addressing the four FAIR Principles has been an open and transparent process involving volunteers glad to collaborate. Every single indicator has been defined and discussed on GitHub issues, since this system allows to track all related comments, proposals and requests. The final document includes 41 indicators to measure the principles of Findability, Accessibility, Interoperability and Re-usability and they are divided into three levels of importance: Essential, Important and Useful. It is important to note that the descriptions of the indicators have a marked high-level perspective to provide flexibility in their implementation. This Working Group output assumes that “Certain indicators may be less important or even irrelevant to some, less data-intensive disciplinary communities. Still, it is essential that different scholarly fields have equal chances to comply with the FAIR indicators” \cite{rda2020fair}. This means that the framework has been provided with a descriptive purpose rather than a prescriptive intention, and therefore the metrics can be implemented with a certain grade of interpretability. Furthermore, the priorities themselves may vary depending on the context of the community or the character of the data repository involved. 

All these characteristics fit well with the goals of the tool we describe in the paper and serve as motivation to use such indicators as our basis. In fact, the FAIR EVA tool has been designed to start working by a generic technical implementation of the specific metrics but is ready to be adapted to particular systems like institutional repositories or disciplinary data portals through plug-ins that balance the metrics, their importance or the indicators for any case. Every indicator can be balanced with specific weights. The default configuration matches the RDA levels of importance (namely, essential, important and useful) with x2, x1.5 and x1 in value to give more weight to more important indicators. Last but not the least, during the process to design, test and iterate our tool we have drawn some suggestions for refinement and improvement for consideration by the teams behind RDA FAIR Data Maturity Model and DSpace development (as our pilot implementation has been upon DIGITAL.CSIC, a DSpace based repository), which we further elaborate in another section of this paper.

\subsection*{FAIR EVA to assist to produce FAIR data}
On the one hand, many scientists do not share their data for many reasons \cite{stall2019make} and those who create mature-enough data to be shared do not usually receive proper credit. On the other hand, institutional repositories tend to have multi-level agendas that include the set-up and maintenance of different types of services to different types of users, and compliance with FAIR Principles is becoming another service. Furthermore, the full adoption of the FAIR Principles requires a set of skills in data management and Information Technologies as well as specific technical implementations in the software solution used by a given repository. In addition, some research disciplines are well acquainted with data-driven techniques, methods and tools in their daily work, whereas others are lagging behind for a full host of reasons \cite{10.18665/sr.316121}. 

Therefore, it may be difficult for some disciplines to be aware of the importance of data reuse and, when they realize it, they often face a significant learning curve. Against this background, one of the objectives of FAIR EVA is not only to measure the current FAIRness level of digital objects located in repositories but also facilitate guidelines for data creators and repository administrators to improve them through feedback and tips. This means that FAIR EVA can mitigate some gaps of knowledge regarding data and metadata quality, supporting data producers to better align with FAIR good practices as well as facilitate data sharing.

\subsection*{Automatic and scalable}
The research data produced every year is constantly growing. Disciplines like Astrophysics or Environmental Sciences produce PBs of data yearly. Only in 2019, ESA's Copernicus program with its Sentinel satellites produced 7.54 PBs of data \cite{copernicusData2019} and this number is increasing every year. Also, many research data collections are dynamic data with regular enhancements and changes, which may prompt the need for new FAIRness assessments. Due to the volume of the research data created every year and the versioning of data collections the evaluation of FAIRness of digital objects requires to be automated, with machine-actionable features that assess certain characteristics without human intervention. Furthermore, a system evaluating FAIR needs to be designed to be responsive to a peak of requests, which means that it needs to be modular and scalable. The tool presented in this work aims to support both characteristics, and its architecture is designed to automatically provide a result based on a modular system. In particular, the interface module is separated from the backend module, which is packaged in a docker image \cite{merkel2014docker} that can be replicated and parallelized when needed.

Thus, the ultimate goal of the system is to provide information on how to improve the research products and digital objects' FAIRness. This means that, in large institutions like CSIC, the number of potential users can be high, and scalability solutions need to be provided. In this case, the application is packaged in a Docker image, that can be easily deployed and pre-configured, so new instances can be created when needed. For example, when a usage peak is detected, a new instance of the processing module can be deployed to increase the system capacity. Furthermore, in order to check that everything works as expected a set of tests are defined and whenever a new version of FAIR EVA is released, everything is checked within a Jenkins Pipeline \cite{Jenkins}. This type of pipelines is oriented to automatize different procedures to ensure the software capacity and reliability, executing different checks in terms of coding style, automatic testing, integration status and licensing \cite{OrvizFernandez2020}.

\subsection*{Factoring repository characteristics and EOSC context}
The FAIR Principles are generic and adaptable to any discipline, but there are some particularities to take into account for some scientific fields. The FAIR indicators provided by the RDA FAIR Maturity Model WG aim to explain what needs to be measured to assess the FAIRness of a digital object and give some tips for implementation. However, while more and more discipline specifics are being taken into account in different initiatives (SSHOC \cite{SSHOPENCLOUD_web}, ELIXIR \cite{ELIXIR_web}, ENVRI \cite{ENVRI_web}...) discussion around existing FAIR enabling technologies in different repository software solutions has not made equal progress. 

Research data and repositories already existed before the FAIR Principles and while some FAIR-related recommendations are strongly internalized in their way of functioning and properties (e.g. minting persistent identifiers to deposited objects, using metadata schemes and vocabularies to describe them in a standard form, promoting access securing such objects are findable and indexed by external databases and service providers through widely used protocols like OAI-PMH, etc.) other FAIR recommendations may turn out to be partially unachievable within certain institutional or scientific disciplines frameworks. For example, different systems for storing research data may have specific characteristics which can not be easily adaptable to particular recommendations of the FAIR Principles. Therefore, in order to be compliant with such principles and associated indicators, institutional repositories need to be capable of adapting them to specific metrics on their reality and context, being realistic and fair with what they can ensure.

In parallel, it is noteworthy to keep in mind that whereas there is a clear promotion of compliance with FAIR Principles within the EOSC ecosystem there is not to date any obligation to score in any specific way or through any specific FAIR assessment tool to be eligible to participate in EOSC as a data provider. There exists a clear framework of EOSC Rules for Participation \cite{eosc_rules} that delineate governance, openness and transparency, research integrity, terms and conditions and technical issues and where FAIR considerations are expressed as follows: “Regarding metrics and certification, it is not expected that any particular threshold of FAIRness is required to be met across all the data in EOSC, where disciplinary norms differ widely. However, the EOSC FAIR principles shall be applied so that a potential data reuser can assess the FAIRness of the data under consideration”\cite{eosc_rules}. Whereas this approach is flexible enough to avoid potential biases, it may also generate a certain degree of uncertainty about what the acceptable minimum FAIR threshold may be. To address this ambiguity the EOSC has put forward the above mentioned task force\cite{fair_taskforce} aiming at implementing the proposed FAIR metrics by assessing their applicability across research communities and testing a range of tools to enable uptake. By promoting different activities such as workshops or hackathons, the "Fair Metrics and Data Quality" Task Force has detected that given the same research product, different tools return widely different results because of independent interpretations of the FAIR principles, indicators or metrics, as well as the data or metadata being gathered. One of the outcomes of this FAIR Metrics Task Force is to propose ways of ensuring that the metadata provision is exactly the same for the different tools. This is essential to obtain similar results across different tools, since they are obtaining the same data and metadata to assess. One of the already available technologies to provide metadata is Signposting \cite{DeSompel2015}, an unambiguous approach to expose the metadata of a repository landing page in a machine-readable manner. This means that the landing page does not need to be interpreted, but the contained information is provided unambiguously. 

Taking into account the different stakeholders and potential users for a given FAIR assessment tools, making tools comparable is very relevant specially for funders. An agnostic and unambiguous way to determine the level of FAIRness is essential, but having an universal and generic way to check across disciplines and applications is really challenging. Apart from obtaining exactly the same data and metadata as the FAIR Metrics Task Force suggests, well-documented details on how the Principles have been interpreted and implemented is necessary. Therefore, FAIR EVA approach suggests to exploit semantics technology to describe the different implemented tests and relationships among Principles, metrics, indicators and even tests run by other tools.

FAIR EVA has been built to measure the FAIRness of digital objects themselves, but also to evaluate how well FAIR is supported in a particular infrastructure. Most data repositories are rapidly evolving and adopting new technologies and features (such as the above mentioned Signposting and ResourceSync), and some consensus has been reached to include generic characteristics, like a set of guiding principles to demonstrate digital repository trustworthiness. In this regard, the TRUST Principles (Transparency, Responsibility, User focus, Sustainability and Technology) \cite{Lin2020} look for a balance between a set of properties and services that will ensure FAIRness and preservation alike. Furthermore, specific characteristics for repositories to go FAIR have been already identified \cite{Hahnel2020} such as the capacity to mint persistent identifiers or offering Application Programming Interfaces (APIs) to enable machine-actionable services. However, institutional repositories are ruled by organizations that may require some restrictions in terms of access to their infrastructure due to  security issues, and therefore enabling  some of such features may be problematic. That is why, although it is recommendable to adopt generic features and interoperable protocols, assessing the performance of FAIR metrics needs to be done by taking into account the specific repository context, so as to provide well-balanced results for its institution or community.

\subsection*{Comparison with other tools}
There have already been several attempts to explore in-depth differences in approaches and methodologies when it comes to execution, measurements and scoring of existing FAIRness tools, thus opening the discussion about the need for raising more awareness on diverging scoring different tools may produce and find ways for closer interoperability and standardization across them. In this regard, one of the earliest attempts was conducted by RDA FAIR Data Maturity Model WG in 2019 to provide a landscaping analysis of existing approaches related to FAIR self-assessment tools (both manual self-assessment and automated tools)\cite{Bahim2019}.
More recently, "A comprehensive comparison of automated FAIRness Evaluation Tools"\cite{Sun2022} has focused on a systematic evaluation of current automated FAIRness evaluators, with concrete suggestions for improving their quality and usability, and highlighting the challenges that these tools pose for their right contextualization, interpretation and adoption. Thus, this paper suggests that given their different coverage, metrics implementation and emphases, transparency and clarity towards potential users should been enhanced to secure wide usability. Furthermore, under the EOSC-Synergy context the Deliverable 3.5 \cite{synergyd35} includes a comparison at technical level among the list of automatic FAIR assessment tools available at that moment. This analysis contributes to propose new ways and solutions to facilitate the matching among different tools, in order to obtain comparable results and smoothen potential biases.

The above mentioned study\cite{Sun2022} identified several differences across tools regarding aspects such as transparency in implementation, user friendliness of scoring display and support documentation after running tests. By way of illustration, an outstanding feature of the FAIR Evaluator tool is its community-driven framework and its scalability to create and publish a new collection of Maturity Indicators (MIs) to meet domain-related and community-defined requirements of being FAIR. Also, whereas FAIR Evaluator publication of its MIs and metric tests in a public Git repository are efforts in favour of transparency in implementation and user understanding FAIR Checker tool \cite{fairChecker} mostly focuses on generating final test results (pass or not pass). Focus degree on specific indicators and metrics may also vary across tools greatly. For instance, it may be stated that F-UJI tool \cite{Devaraju2021} has comprehensive metrics for Reusability, while the FAIR Evaluator focuses on the Interoperability. F-UJI requires that the relationship properties that specify the relation between data and its related entities have to be explicit in the metadata and uses pre-defined metadata schemas (e.g., “RelatedIdentifier” and “RelationType” in DataCite Metadata Schema). Compared to F-UJI, the FAIR Evaluator has a broader requirement for acceptable qualified relationship properties by including numerous ontologies which include richer relationships. 

For its part, the EOSC synergy deliverable 3.5 compares the set of indicators implemented by automatic tools, mapping the tests with specific FAIR principles to define which of them can be compared. Table \ref{tab:compare} is included in the deliverable and based on the Principles and subprinciples, it exposes which of them are addressed by the list of automatic tools. Furthermore, the Annex I from the deliverable details different aspects about the tools, like the code availability, running services, license, type of reporting, type of outputs and the community behind the tools.

\begin{table}[]
\centering
\begin{tabular}{c|c|c|c|c|c|c|c|c|c|c|c|c|c|c|c}
          & F1 & F2 & F3 & F4 & A1 & A1.1 & A1.2 & A2 & I1 & I2 & I3 & R1 & R1.1 & R1.2 & R1.3 \\
\hline
F.EVA     & Y  & Y  & Y  & Y  & Y  & Y    & Y    & Y  & Y  & Y  & Y  & Y  & Y    & Y    & Y    \\
F-UJI     & Y  & Y  & Y  & Y  & Y  & N    & N    & N  & N  & Y  & Y  & Y  & Y    & Y    & Y    \\
FES       & Y  & Y  & Y  & Y  & N  & Y    & Y    & Y  & Y  & Y  & Y  & N  & Y    & N    & N    \\
F.Enough  & Y  & Y  & Y  & Y  & N  & Y    & Y    & Y  & Y  & Y  & Y  & N  & Y    & N    & N    \\
F.Checker & Y  & Y  & N  & N  & N  & Y    & N    & N  & Y  & Y  & Y  & N  & Y    & Y    & Y   
\end{tabular}
\caption{Overview of the FAIR Principles being assessed by five different automated FAIR assessment tools \cite{synergyd35}}
\label{tab:compare}
\end{table}

\subsection*{FAIR EVA contribution}
Against this background, FAIR EVA was conceived as an automated tool with the value added service to be able to run tests and score FAIR Principles compliance through a predefined system of plug-ins. The vision was that every single plug-in may be customized as needed, for example, to support specific FAIR metrics/indicators and/or specific technical features of data repositories. From the outset it was clear that the one size-fits-all approach would not help everyone within the varied community of data infrastructures in order to measure and improve their FAIRness and therefore, the idea to develop a tool that could be easily tailored to the specifics of each of such data infrastructure seemed a promising path. Although this community-driven approach has been adopted by other tools like the FAIR Evaluator, the innovative contribution of FAIR EVA lies in its plug-in-based architecture, which is capable of adapting to different ways of data and metadata gathering. In this context, it was decided to start building the tool through a pilot implementation on top of DIGITAL.CSIC multidisciplinary repository, given the presence of a number of FAIR supporting functionalities on its infrastructure (e.g. open and standardised interoperability protocols, a large collection of open access data objects with PIDs, use of metadata standards and some controlled vocabularies etcetera). In addition, it was deemed a good opportunity to explore in full detail specific features in multidisciplinary data repositories when the few previous automated tools tended to rather focus on domain specific repositories/communities. 
Last but not the least, another distinctive feature of FAIR EVA is its capability to adjust to a given repository implementation, that is to say, by taking into account its specific metadata guidelines and recommendations, policies and advanced functionalities when running the tests: indeed, the underlying incentive of this pilot project was to build a new institutional service on top of an existing  repository as it is of little use to have a tool that a priori demands the existence of this or that technical/metadata property if such property is not available on the repository infrastructure as a result of its own nature, policies and/or software. It is this scalable and modular nature, on the one hand, and this specific repository service oriented approach, on the other hand, what we find as an additional contribution to the landscape of existing automated tools.

In summary, FAIR EVA proposes a set of advances from the state of the art, providing a flexible system to adapt to different contexts. First of all, the architecture of FAIR EVA enables a scalable solution to assess several digital objects or datasets, adapting to the growing volume of research products. Secondly, FAIR EVA is intended to be adaptable thanks to its plug-in system, which allows the customization of connection to different types of data systems or repositories as well as the re-implementation of base metrics, the redefinition of indicators, the extension of tests, the configuration of specific data characteristics or metadata terms and the definition of weights in the metrics, thus taking into account the reality of different scientific domains and multidisciplinary repositories. Further, this customization is extensible to feedback to users, so that any data system can tailor the information provided to their specific users. Thirdly, the objective of FAIR EVA as an advisor is to guide data producers to improve the quality of their data as well as support repository managers to improve the service provided. Finally, FAIR EVA makes available technical solutions that have been developed as ontologies and semantics to facilitate the comparisons with other existing automated tools. These features are not addressed by any other tool, and are relevant to adapt to modern data contexts like the EOSC, where the Open Science paradigm is adapted and the amount of research products are growing exponentially.

\section*{Methods}

\subsection*{Taking into account the institutional environment: No repository is an island}
FAIR EVA is a development that falls under EOSC-Synergy \cite{synergy_web} agenda to support FAIRness in participating data repositories including DIGITAL.CSIC as well as thematic services representing different scientific disciplines and data systems. Likewise, this tool nicely fits within DIGITAL.CSIC strategy around research data management: the repository collection policy included research data amongst accepted research outputs in 2010 and there has been a gradual adoption of related value-added services ever since. This path is common amongst institutional repositories that have broadened their focus from management of peer-reviewed publications solely to a fuller spectrum of institutional outputs. In such transformation, this new type of output joins an infrastructure designed from its inception to manage works from different disciplines and with a primary focus on providing their access. This legacy mission determines the choice of agnostic metadata schemes such as DCMI and DataCite, which are indeed the most widely used metadata formats used by data repositories according to Re3data \cite{re3data} (with 433 repositories reporting to use Dublin Core and 262 using DataCite at time of writing this paper). Another legacy characteristic revolves around the emphasis of such repositories on labelling their outputs concerning compliance with different types of open access policies, both institutional and third parties. Last but not least, repository software legacy issues are also key to keep in mind, given that as a rule of practice most institutional repositories are projects that have been born by the initiative of the library community within their institutions and given the long record of budget and human resources constraints faced by them a common approach has been to maximize on the infrastructure and resources already put in place. DIGITAL.CSIC was first installed on DSpace in 2008 and is currently running on DSpace CRIS v5.10.

Amongst data services, DIGITAL.CSIC joined DataCite in 2016, which has allowed DOI assignation to over 12,000 research data and other non-traditional outputs (such as research software, and preprints). In parallel, DIGITAL.CSIC team launched a training programme around research data management in 2015 for the institutional community and the repository received the "Data Seal of Approval" certification at the closing of 2015. Awareness raising and support to the institutional community have also been channelled through other means, notably through the upskilling of CSIC libraries professionals; the elaboration of guidelines and good practices; and the participation in several initiatives that promote the upgrade of data as primary research outputs. Research data services in DIGITAL.CSIC have been on the rise over the last years due to the multiplication of data related funders and publishers policies. Demands to use the institutional repository as data provider in some EOSC thematic infrastructures have also contributed to paying further attention to other aspects around research data management.

Further, CSIC issued its institutional Open Access mandate on April 1, 2019 \cite{Mandato}, exactly 14 years after the institutional signing of the Berlin Declaration. This mandate falls under the so-called green mandates as it requires institutional researchers to deposit the metadata of their peer-reviewed publications and underlying datasets into DIGITAL.CSIC. As far as research data are concerned, the institutional mandate explicitly refers to data that supplement peer-reviewed publications and emphasizes that data have to be FAIR. Annual monitoring about compliance with institutional mandate\cite{mandatoMonitor}shows high rates of performance as regards peer-reviewed publications obligations however research data-related policy requirements are still far from full alignment.

FAIR EVA comes to enrich this state of things by providing both DIGITAL.CSIC administrators and data creators with a tool to assess the degree of alignment with FAIR principles recommendations. On the one hand, the FAIR EVA is a powerful tool for repository administrators to easily identify those FAIR indicators yielding lower scores and prioritize them in the repository technological and training agenda. On the other hand, the tool provides data creators with clear tips to improve the description of their data.

Therefore, FAIR EVA can bring the following benefits to different constituencies actively engaged in research data management:

\begin{enumerate}
\item For repository’s administrators
\begin{itemize}
\item Identification of pending technical integrations to make FAIR Principles across all collections in the repository 
\item Identification of aspects in the description of a specific dataset that require further attention by the administrators and/or data creators
\item A tool to analyze differences in the degree of compliance with the FAIR Principles and indicators across research domains. The possibility to compare these differences are quite useful for multidisciplinary data repositories as they face the major challenge of securing common functionalities for the management of digital contents while offering incremental support for specific features related to disciplines
\item A practical way to market and showcase the infrastructure as a valid and trustworthy amongst its community of users
\item A tool to refine repository’s policies and improve the alignment of the repository with all relevant recommendations for research data description and data repository management 
\end{itemize}
\item For data creators/researchers
\begin{itemize}
\item A quick way to find out whether their datasets are documented sufficiently as to enable findability, accessibility, interoperability and reusablity and what to improve
\item An easy to use tool that help data creators raise their awareness about the importance of their data to be FAIR
\item A quick way to find out what support a given repository is offering to make FAIR Principles work in practice
\item An evidence based assessment that can be attached to projects final reports and other related deliverables
\end{itemize}
\item For data creators institutions, funders and publishers
\begin{itemize}
\item An evidence based assessment about the degree of compliance with their research data policies and/or mandates
\item A tool to help them decide whether a specific collection of datasets is FAIR enough to be referenced and showcased to the broader community
\end{itemize}

\item For repository software developers
\begin{itemize}
\item A tool to identify potential developments on the code for closer alignment with FAIR Principles
\item A practical way to further engage with the community behind the promotion and realization of FAIR indicators
\end{itemize}
\end{enumerate}

\subsection*{RDA Indicators implementation in FAIR EVA}

As above mentioned, RDA FAIR Data Maturity Indicators were selected to automate the scoring of FAIR Principles compliance in FAIR EVA tool given their broad acceptance in the global community, their collaborative and transparent nature and flexibility to implement. These indicators have been implemented taking into account the context of DIGITAL.CSIC repository as first running example.

As already explained, the goal of FAIR EVA is not only to provide a rating or percentage of FAIRness of a digital object  but also guide  researchers and repository administrators to improve compliance. This is done by explaining the problems found for any specific metric during the assessment, and by linking to a particular training or providing details on what is missing and what can be added to improve scoring. The user perception is very relevant, and feedback mechanisms are enabled to establish direct communication. The Table \ref{table:rda_in} includes a list of selected RDA indicators, how they are implemented and what feedback is generated together with a link to the full details in the documentation \cite{FAIR_eva_tech}. Further, FAIR EVA tool proposes a layering system, based on a generic implementation that can be extended for a particular context via plug-ins. A plug-in can include discipline and/or policy -specific information like metadata terms to check but also can redefine the technical implementation methods to ensure that a specific metric takes into account repository and/or disciplinary characteristics. For example, a potential redefinition may focus on the way data and metadata in a particular data service are accessed or let the data service administrators calibrate the weights of every single indicator or metrics considering the nature/context of the service. This flexibility helps get an overall and meaningful feedback. By default in FAIR EVA, the weights are defined according to the level of importance of RDA indicators, that is to say, x2 for essential, x1.5 for important and x1 for useful indicators. The final value is obtained from the following formula, being \textit{Ts} total score, \textit{Pi} number of points of indicator \textit{i} and \textit{W} weight of indicator \textit{i}.

\begin{equation}
{Ts = \frac{\sum(P_{i} x W_{i})}{\sum W_{i}}}
\label{for:score}
\end{equation}

As a real test runs, feedback is provided whenever the score of a given indicator does not reach 100\%. Feedback distinguishes between technological errors that refer to repository administrators and metadata incompleteness where data creators are encouraged to provide a richer description and/or manage data files according to the repository policies and recommendations. However the tool has been built having research data in mind as the primary focus, it is also being promoted to assess the FAIRness degree of any other type of research output.

\begin{table}[h]
        \centering
        \small
        \setlength\tabcolsep{2pt}
        \begin{tabular}{|c|p{7cm}|p{8cm}|}
\toprule 
RDA Indicator & Technical implementation & Feedback \\
\hline
RDA-F1-01M & Searches within a predefined list of potential metadata terms to identify the metadata (dc.identifier.uri and dc.identifier.doi) if any information is available. & All items deposited in DIGITAL.CSIC are granted a handle by default. This PID is packaged in dc.identifier.uri. DOI is minted by DIGITAL.CSIC for all datasets that do not have a DOI already, if your dataset has not been given one, please contact DIGITAL.CSIC Technical Office \\
\hline
RDA-F1-01D & Searches within a predefined list of potential metadata terms (dc.identifier.uri, dc.identifier.doi and dc.relation.publisherversion) to identify the data if any information is available. & If the DOI/Handle has not been generated, contact repository administrators. If the digital object had another DOI/PID before deposit on DIGITAL.CSIC, add the dc.relation.publisherversion in metadata registry to point the external site where the data files are hosted. \\
\hline
RDA-A1-02M & Looks for the metadata terms in HTML in order to know if they can be accessed manually & If metadata are not available through the landing page of the digital object, please report to DIGITAL.CSIC. \\
\hline
RDA-A1-02D & Checks the presence of an access metadata term. In DIGITAL.CSIC this information is packaged in dc.rights. The metadata element dc.description is also used to provide extra information whenever data are not accessible on DIGITAL.CSIC.
 & If data files are not available open access in DIGITAL.CSIC or any other site please indicate how they could be requested as a private copy in the metadata element dc.description. Please provide with any other relevant information to facilitate access. \\
\hline
RDA-I1-02M & Checks, via OAI-PMH, if the metadata can be retrieved in a format like RDF & OAI-PMH supports RDF \\
\hline
RDA-R1-01M & Depending on the metadata schema used, checks that at least the mandatory terms are filled (75\%) and the number of terms are high (25\%) & It is good practice to describe digital objects as richly as possible to ease reuse. Please check DIGITAL.CSIC guide for full details and pick up the template of the resource type you are describing to make sure that you follow all DIGITAL.CSIC metadata and supporting files recommendations. \\
\hline
RDA-R1.1-01M & Checks if the license information is available in any format. In DIGITAL.CSIC this information is packaged in dc.rights.license & Don`t forget to include the standard usage license of the digital object in the dc.rights.license metadata. It is highly recommended to insert it in an URL format. The following tool (https://ufal.github.io/public-license-selector/) helps identify the most appropriate standard license for datasets and software. \\
\bottomrule
\end{tabular}
\caption{Snapshot of the Template of technical implementation and user feedback created for DIGITAL.CSIC plug-in}
\label{table:rda_in}
\end{table}

\subsection*{FAIR EVA Architecture}
\begin{figure}[h]
\centering
\includegraphics[keepaspectratio=true,scale=0.3]{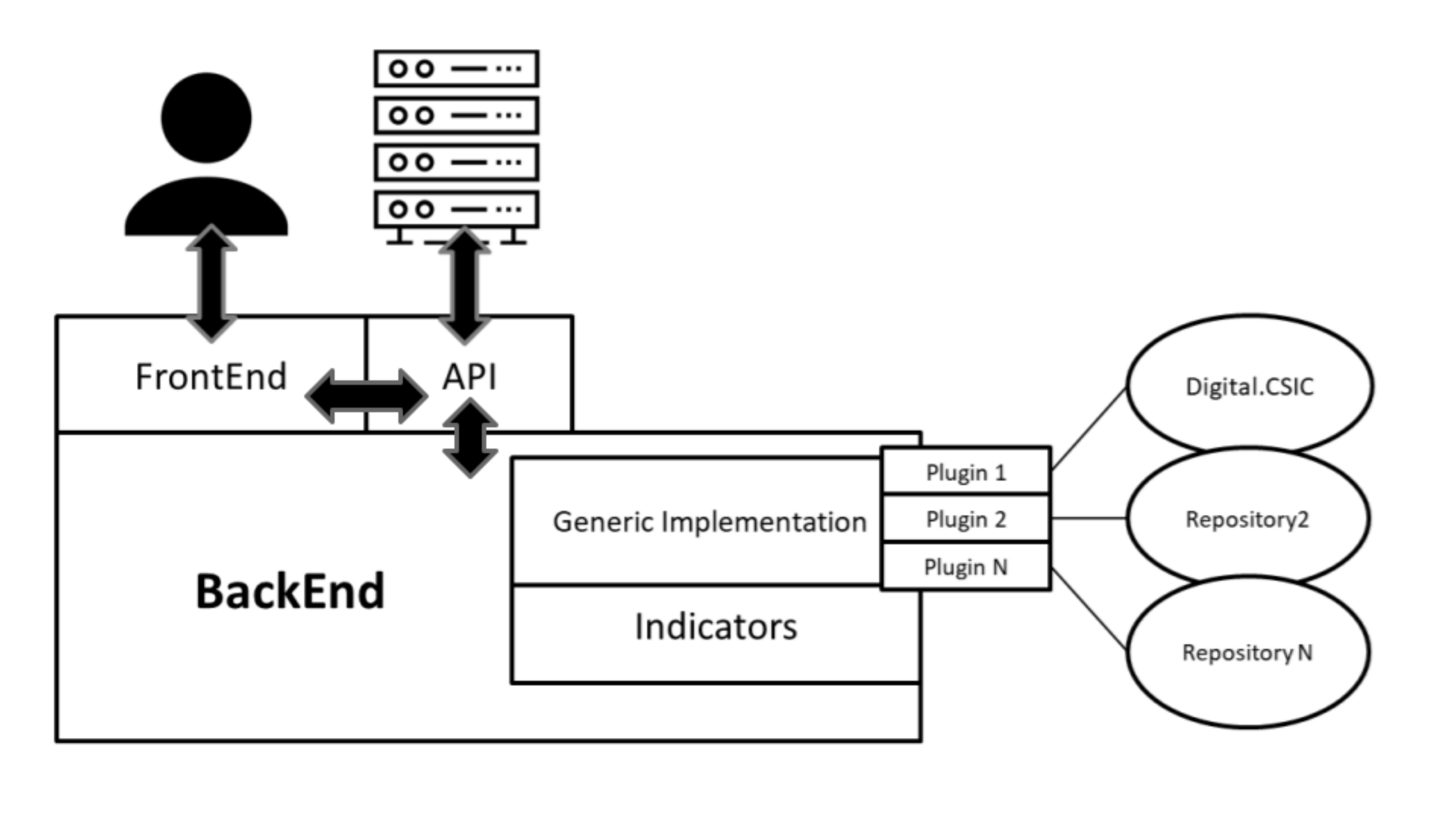}
\caption{FAIR EVA Architecture}
\label{fig:arch}
\end{figure}

The approach adopted to address metrics performance aims to be modular, dynamic and customized. This means that the service includes two basic layers developed in Python: the back-end implements and performs all the tests defined and the front-end layer allows the user to access and execute the tests in a web interface or an Application Programming Interface (see Figure \ref{fig:arch}). To make the application easily scalable, any component is configured as a service and packaged in a Docker image.

\subsubsection*{Front-end and API}
The upper layer in the architecture, which connects to the user, is a web application as well as an API that calls the proper methods to perform the metrics. The web interface is based on Flask Python library \cite{flask} that renders a set of templates to let the user manage the configuration of the test. The application receives the identifier of the resource to be checked in terms of FAIRness and selects the proper repository or data service to ask for. It can be also configured to work with just one specific data service. To make the system reliable and robust, the web application calls transparently and directly the API, which is defined and developed in a standard way thanks to OpenAPI specification \cite{openApi}. The available indicators, methods, accepted requests and provided outputs are defined in a YAML file (fair-api.yaml). The default version includes all the indicators proposed by the RDA, but they can be extended to implement new specifications. The YAML file defines also how the API needs to be called and what information can be included in the response. In this case, the indicator methods provide a JSON block including the name of the indicator request, a percentage about how well this indicator has been addressed and a feedback message including information about how this metric can be improved by the user. All this information can be adapted and customized by the repository administrator, in order to provide concrete feedback information and guidance. Furthermore, within the YAML different attributes can be configured, like the weight or importance of the specific metric to be adapted to the repository or the discipline needs. This is important since the user received an overall result of how well the FAIR Principles are addressing the specific repository criteria. Thanks to this hybrid approach web interface and API integration, the system can support users manually but also machine-actionable features to automatize and scale the requests.

\begin{figure}[h]
\centering
\includegraphics[keepaspectratio=true,scale=0.3]{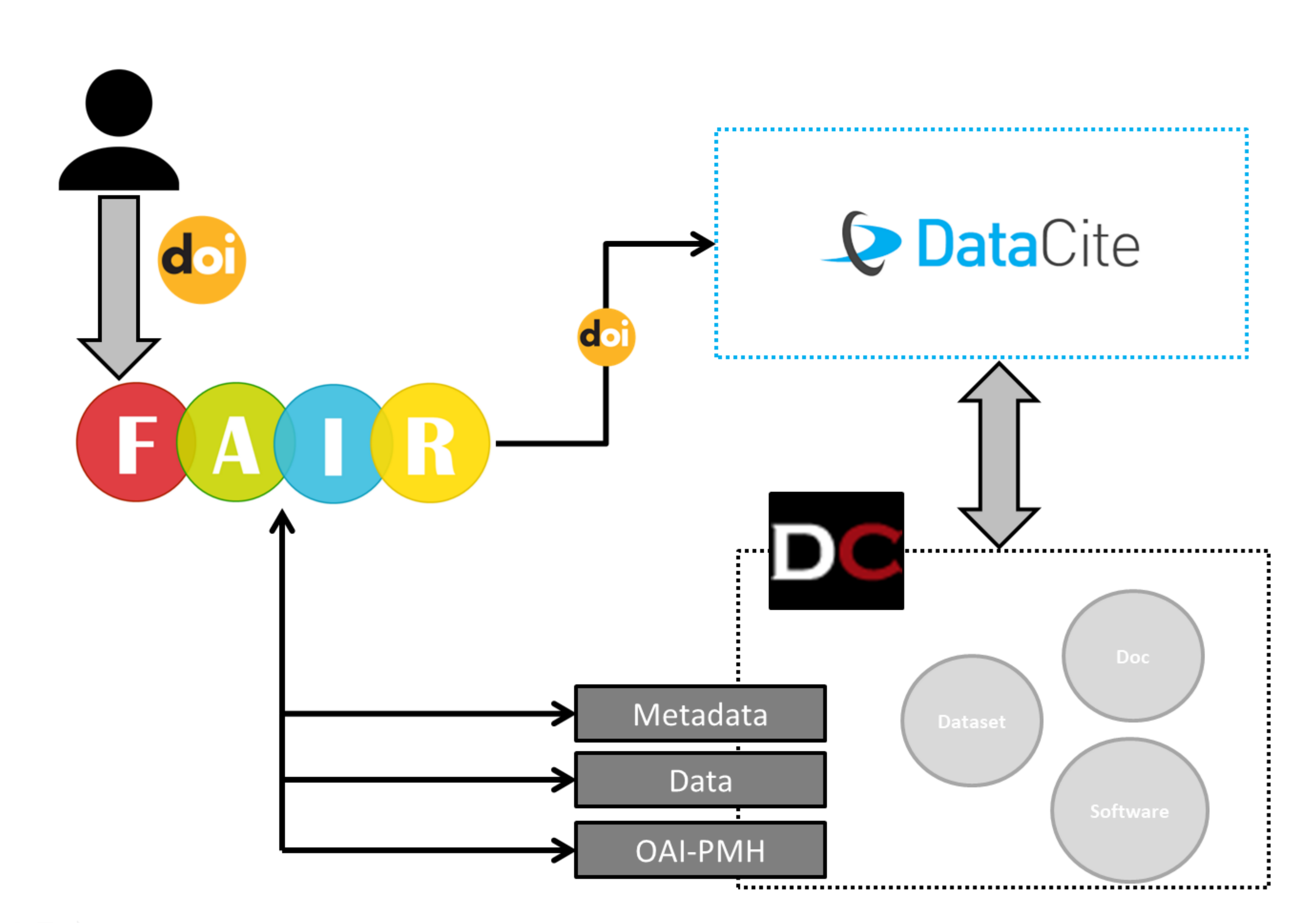}
\caption{DIGITAL.CSIC Plug-in - Flow description}
\label{fig:flow}
\end{figure}

\subsubsection*{Back-end}
The application back-end implements all the indicators/metrics defined in the YAML file from the Front-end. It takes advantage of the Python inheritance features, which allows creating a Class including attributes and methods that can be extended in child Classes. The system adopts a “Plug-in” approach, so for each set of indicators, a new class implementing the metrics is defined. This parent class generically implements the different tests, capable to be executed but without taking into account repository-specific characteristics. To provide more details, a new Plug-in for a concrete repository can be created, which will be a child Class from the parent one. This child Class extends the set of attributes and methods, which can be re-implemented or extended or even keep the original actions if no new definition is needed. An example plug-in is included in the code repository to help developers to create a new one. Since Python allows multiple inheritances, a single repository plug-in can extend multiple indicators classes, that can be easily concrete. For example, the RDA indicators are defined in the Front-End YAML file, which is implemented generically in an RDA Class. This class is extended in a child Class called DigitalCSIC, which has the same attributes and methods as the parent one. Some of the methods are redefined in order to specify how the RDA indicators need to be executed specifically on DIGITAL.CSIC institutional repository. The flow that executes this Plug-in can be found in Figure \ref{fig:flow} .

\subsection*{Workflow for managers}
\begin{figure}[h]
\centering
\includegraphics[keepaspectratio=true,scale=0.8]{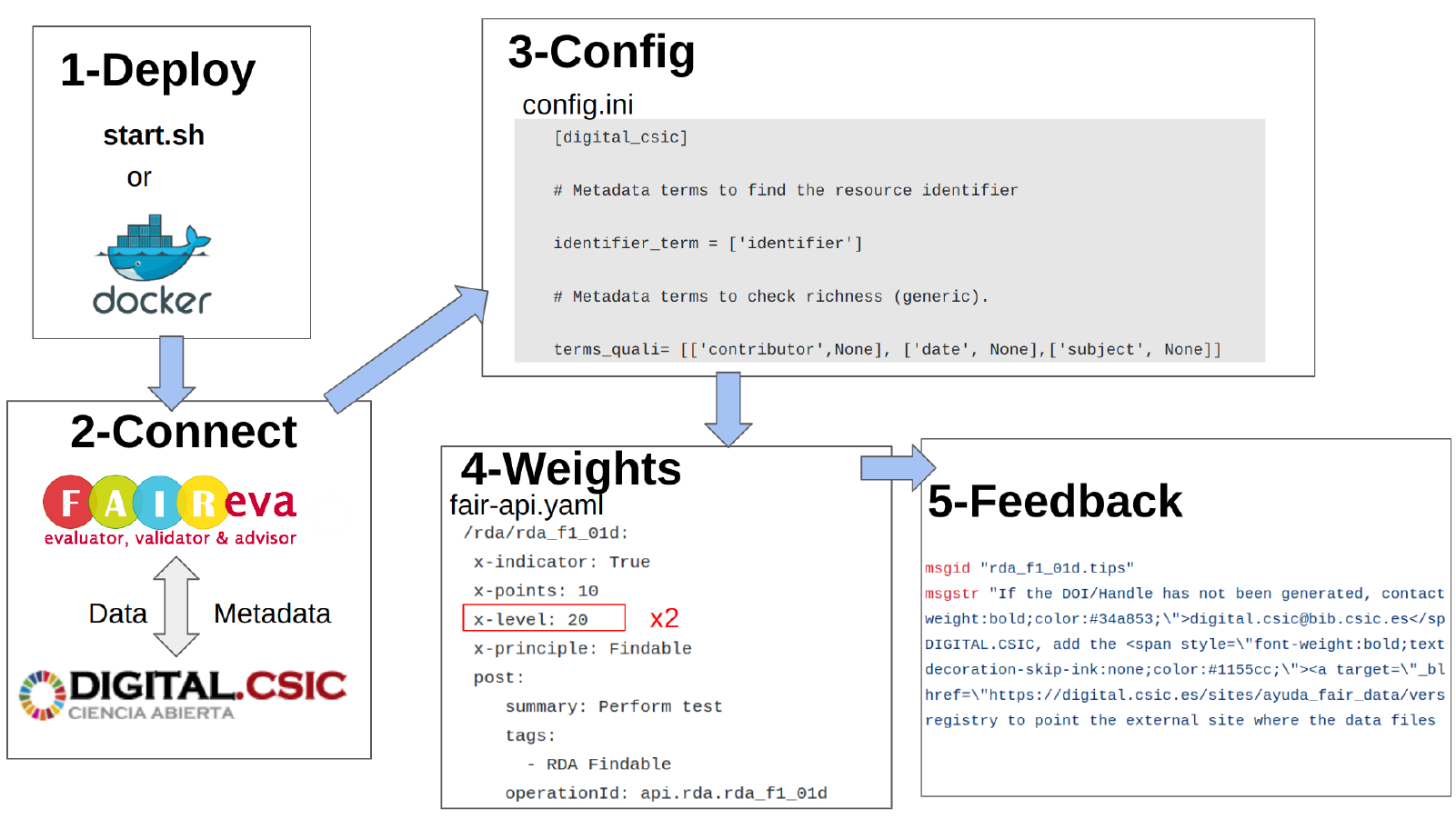}
\caption{Workflow for deploy and config a FAIR EVA instance}
\label{fig:manager_flow}
\end{figure}

FAIR EVA acts as a stand-alone service connected to the data repository or service. All documentation related to its operation can be found in the code repository \cite{fairEvaRepo} and the workflow for the deployment is described in Figure \ref{fig:manager_flow}. FAIR EVA can be deployed using the "start.sh" executable or through the deployment of an available Docker image. Once the service is active, the repository manager must connect it to their system. To do this, the generic OAI-PMH based plug-in can be used or a customized one can be developed. 
In the case of DIGITAL.CSIC, a specific connection plug-in has been developed. Each plug-in defines the way in which data and metadata are accessed and inherits or extends each of the indicator-based tests, which can be redefined. Once the plug-in is successfully connected, in the specific section of the configuration file "config.ini" of the plug-in used, it is specified which metadata terms are parsed. For example, the configuration item "identifier\_term" indicates which metadata terms may contain the identifier of the digital object. Finally, the weights for each of the flags can be redefined in the configuration file "fair-api.yaml", which by default respect the value of the RDA flags. Furthermore, the feedback provided for the data producers can be customized under the “translations” folder. Every RDA indicator is identified by its code (e.g. "rda\_i3\_03m") and the feedback provided for the user can be edited in the proper config line (e.g. "rda\_i3\_03m.tips") in different languages.
In order to evaluate a number of digital objects, different FAIR EVA API requests can be programmed automatically.

\subsection*{Workflow for users}
Once the FAIR EVA web service is deployed and configured with a connection to a repository, users can evaluate their deposited digital objects. The user can access the endpoint of the service and simply insert the identifier (either DOI or Handle in the case of FAIR EVA plug-in for DIGITAL.CSIC) of the digital object to be evaluated. The tool then accesses the metadata of the digital object and, if any, associated data files via DSpace API and directly from the record page in the repository infrastructure and an automated assessment is generated within few seconds.

\begin{figure}[h]
\centering
\includegraphics[width=\linewidth]{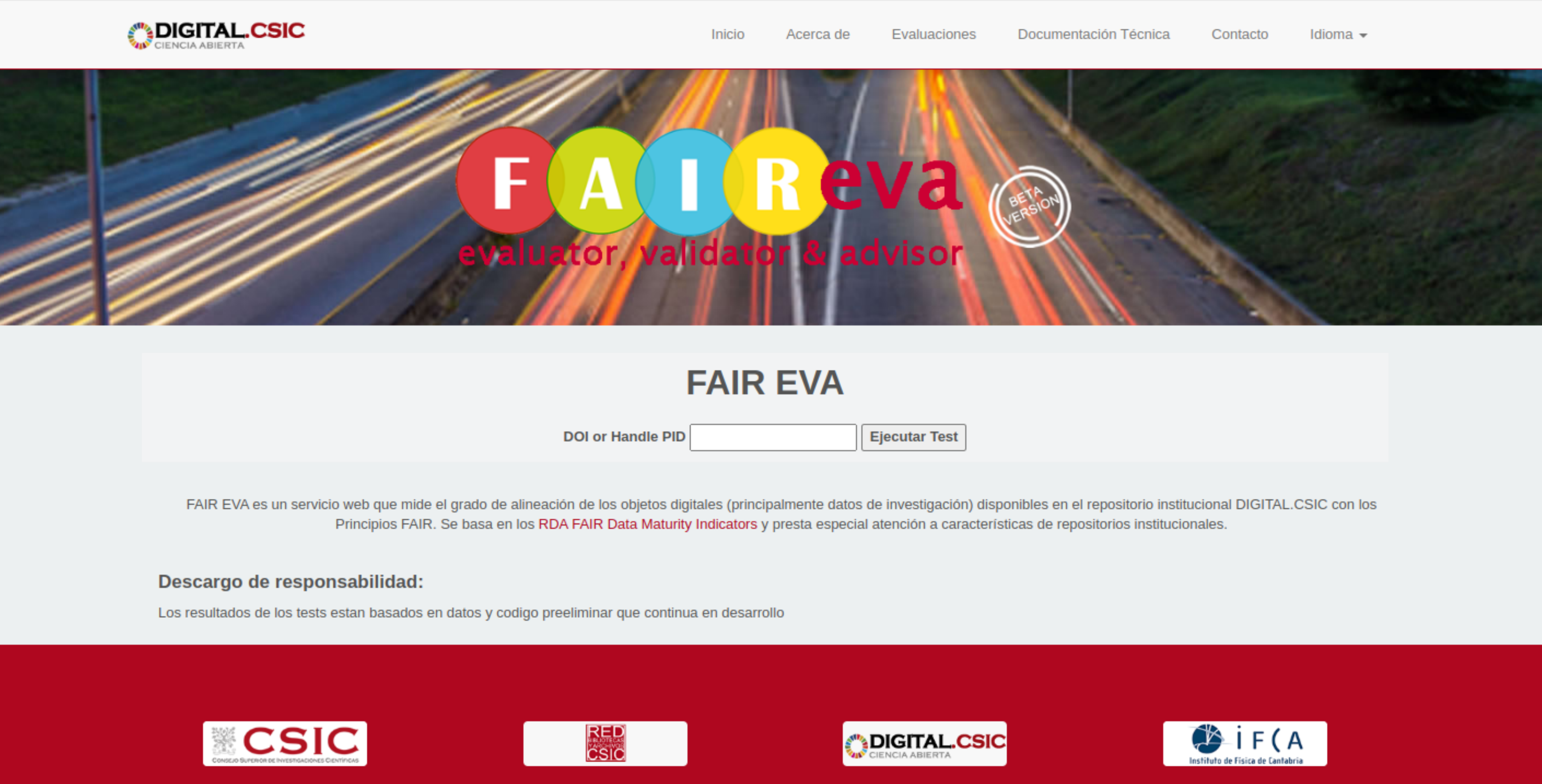}
\caption{Homepage of the web service https://fair.csic.es/es}
\label{fig:user_view}
\end{figure}

\begin{figure}[h]
\centering
\includegraphics[keepaspectratio=true,scale=0.3]{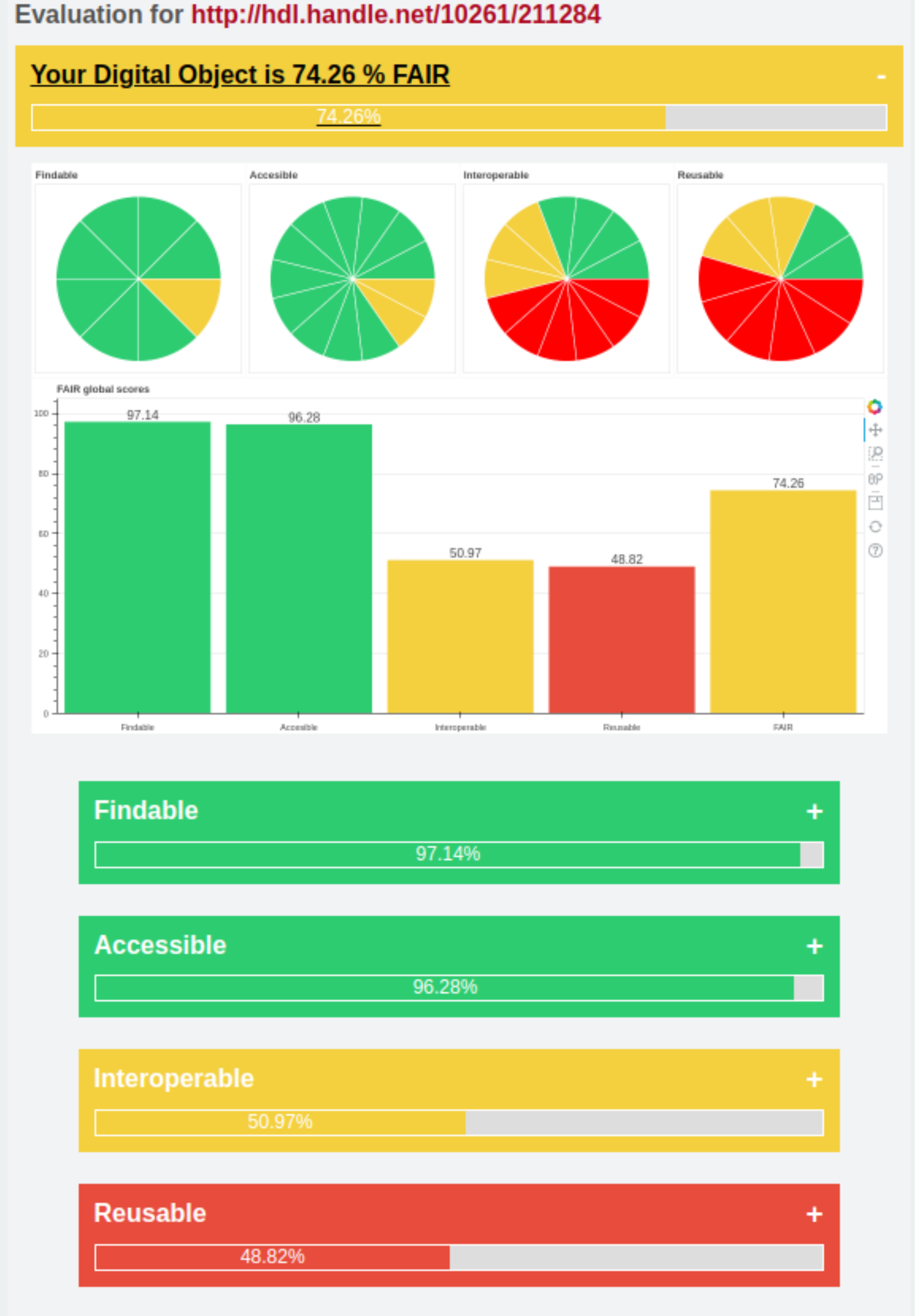}
\caption{Look and feel of the overall assessment of the dataset at https://digital.csic.es/handle/10261/211284}
\label{fig:results_view}
\end{figure}

After running the analysis, the user is taken to the evaluation page, where she receives a multilevel assessment of the FAIR level of her digital object, with metrics for each test (Figure \ref{fig:results_view}). The assessment generates an overall score in percentage format, and immediately below the user can see the overall score as distributed across each of the 4 sets of indicators (grouped by Findable, Accessible, Interoperable and Reusable). Each group of indicators gives the user the possibility of accessing more granular assessment for every single RDA indicator taken into account to estimate the overall score for its group. Thus, for each test, a drop-down menu shows information for every single RDA indicator, namely, its name;  the "Indicator Level" (Essential, Important, Useful), which is connected to its specific score weight; the indicator description according to RDA ("Indicator Assessment"); the specific technical implementation in the plug-in ("Technical Implementation"); outcome of the test execution at technical level ("Technical feedback"), and recommendations for improving FAIRness  ("Tips"). An example of the structure of the assessment reports can be seen in the below (Figure \ref{fig:results_view}). In addition, the user can export the full assessment report in PDF format for her own convenience.

The full assessment report provides with rich information that is useful for repository administrators, data creators and data depositors. Repository administrators obtain a full picture on how every single RDA indicator scores in the infrastructure as well as details on the FAIRness degree of specific datasets. The assessment sections about technical feedback and tips are particularly useful for users willing to improve the score of their digital object. Further, in an attempt to make all this information more understandable for users and encourage action, the repository team has created a support web \cite{ayuda_fair_eva} with practical information about the metadata that have not scored well enough in the test and how to best use them in the repository. This support web includes explanations for every single metadata used in the repository and taken into account by FAIR EVA as well as good practice examples and links to the repository user guides. The set of tips for DIGITAL.CSIC can be found in the code repository \cite{fairEvaRepo} and the below figure is an example of how the assessment section “Tips” assessment section redirects the user to the support web.

\begin{figure}[h]
\centering
\includegraphics[keepaspectratio=true,scale=0.3]{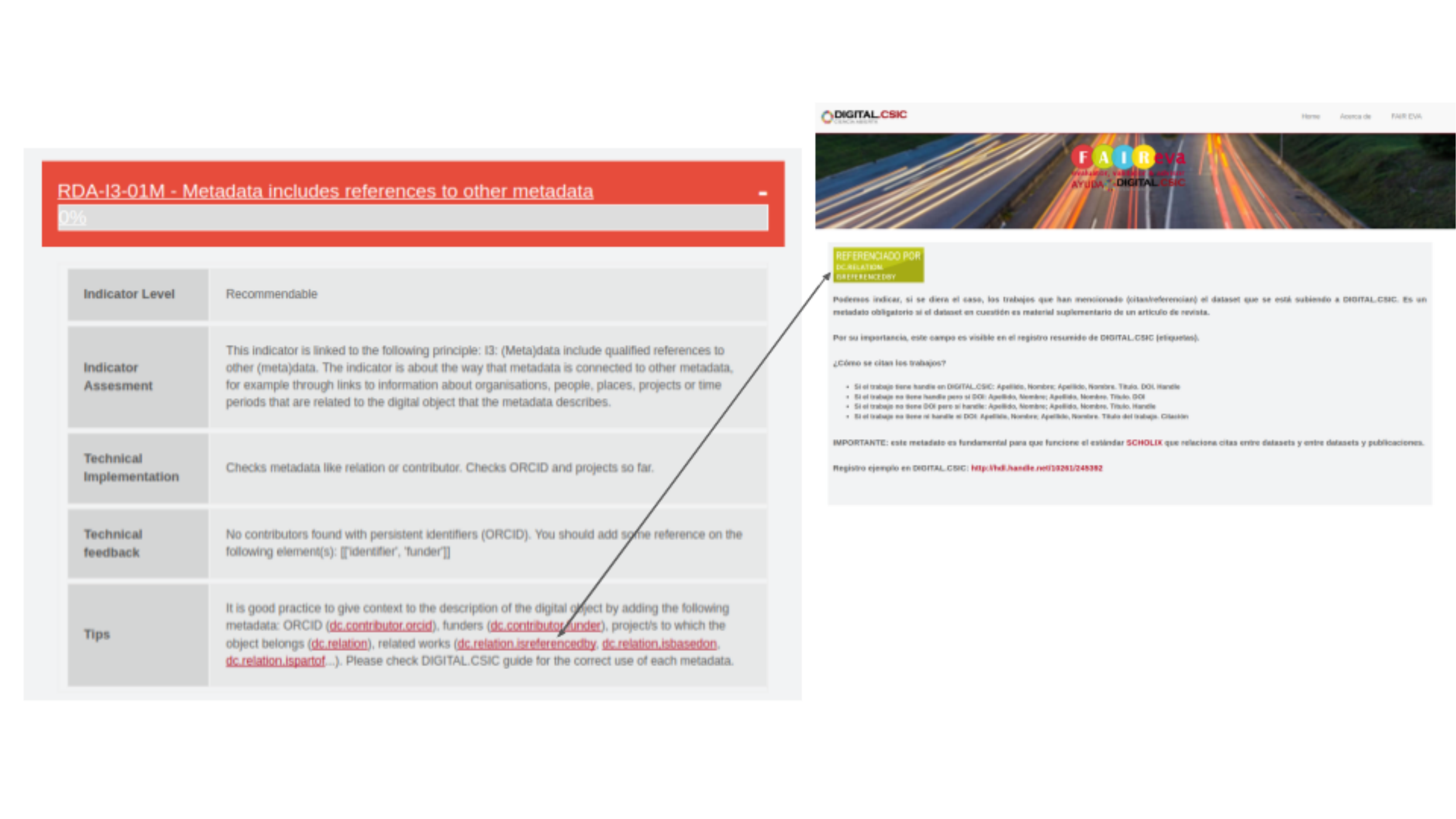}
\caption{Snapshot of the assessment of one RDA indicator in a real example and how the Tips provided lead users to the support web site}
\label{fig:indicator_help}
\end{figure}

In parallel, in order to raise awareness about this new repository service and promote usage, a FAIR EVA icon has been painted in all datasets pages on the repository that redirects users to the main page of FAIR EVA web service.  Repository team started to promote this new service actively amongst institutional data creators and librarians last Fall, through hands-on training and one-to-one support. Preliminary feedback from users has been overall positive, with growing cases whereby assessed datasets have been improved after running a test. However, the repository team has already been asked to run further dedicated training on the tool so that data creators and support staff such as librarians fully understand all possible actions thereafter such as planning systematic reviews of selected dataset collections on the repository or including the assessment reports in deliverables to funders. In addition, the repository team is already working on technological improvements in the infrastructure to better align with some indicators which fall under their scope, such as enabling Signposting and integrating new controlled vocabularies.

\begin{figure}[h]
\centering
\includegraphics[keepaspectratio=true,scale=0.3]{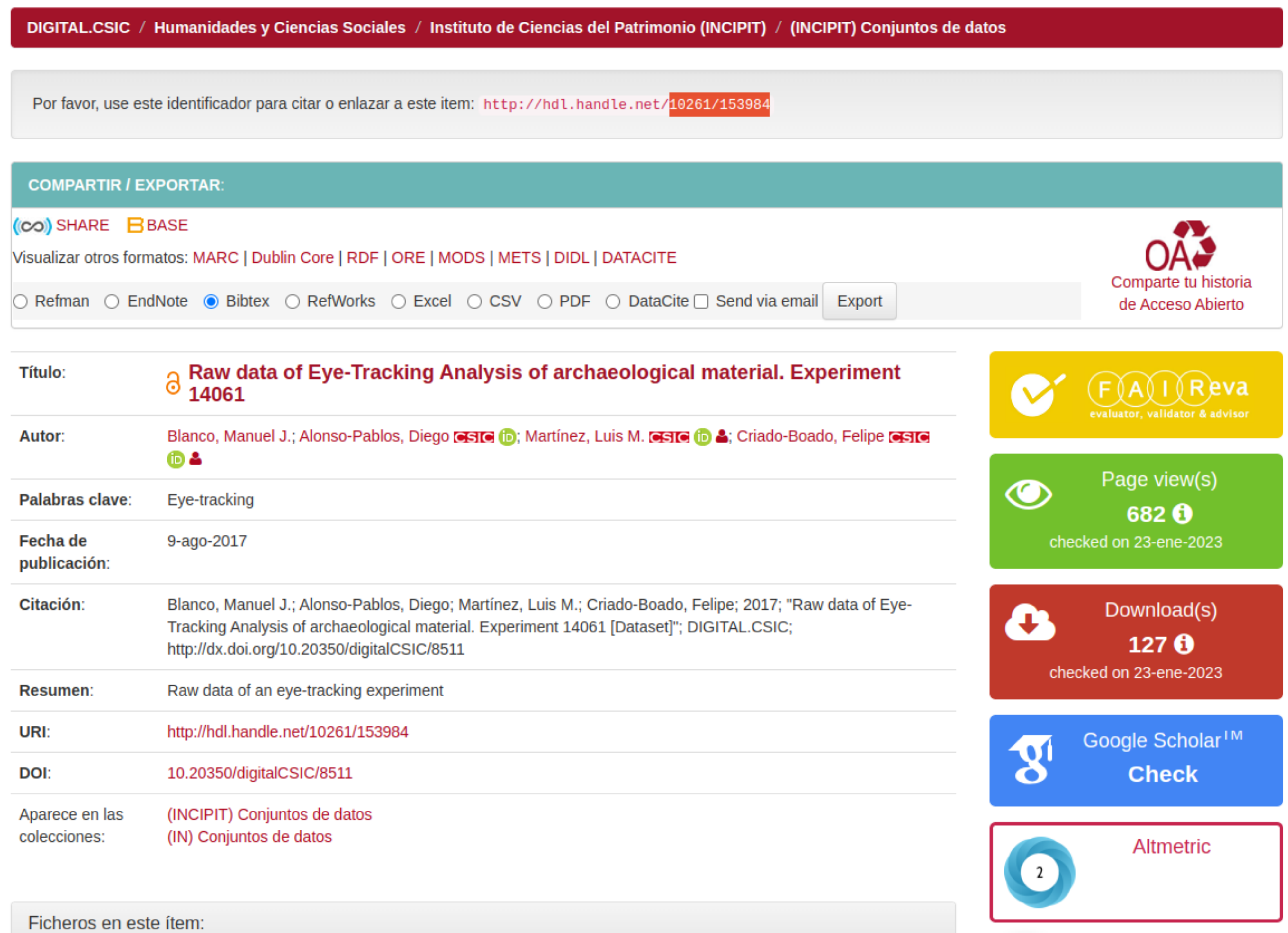}
\caption{Look and feel of a dataset record page at DIGITAL.CSIC promoting the usage of FAIR EVA tool https://digital.csic.es/handle/10261/153984}
\label{fig:landing_page}
\end{figure}

\section*{Results}

After developing the service and the generic plug-in, a specific implementation has been done for DIGITAL.CSIC. As a multidisciplinary repository, it hosts datasets from different scientific domains with varying degrees of compliance with FAIR principles.

Before running the tests, we had two assumptions: on the one hand, the first 2 sets of FAIR RDA indicators (those regarding findability and accessibility) mostly deal with basic technical specifications to support that hosted contents can be discovered, located and accessed on the web: thus, the responsibility to build a public repository where data collections can be found; enable standard metadata schemes to describe the data; allow for human and machine readability of contents; assign persistent identifiers to data and metadata; elaborate contingency plans so that metadata will remain accessible in the long term even if dataset files are not and implement other standard web communication protocols lies on a group of professionals that are not the data creators. On the other hand, however the following 2 sets of FAIR indicators (those regarding interoperability and reusability) still include technical considerations that fall under the repository's administrators' agenda (e.g. integration of standard vocabularies and/or ontologies, interoperability with other open infrastructures, mappings with other metadata schemes, etc.) other aspects do greatly depend on the granularity of the descriptions of the research data. It is in this latter set of aspects where the contribution of data creators may be the greatest as it is often the case that however the infrastructure of data repositories provide standard metadata forms and controlled vocabularies and recommendations to manage research data interoperability and reusability principles are not always completely maximized due to poor metadata description and lack of supporting documentation provided by data creators.

Our pilot assessments have shed light on the above mentioned assumptions and seem to support the principle that while a given repository should always fare the same as far as infrastructure-related indicators are concerned, variance can be found regarding metadata and data conformity. Such variance can be explained for a host of reasons, including lack of sufficient awareness amongst data creators about the importance of metadata to meet the FAIR Principles; absence of community/standard ontologies in a given discipline; or unknown information about the context of the resource being described. It is also important to consider whether the resource being described in the repository is part of a community-driven effort whereby a consensus has been agreed amongst all data creators and stewards and also whether data creators have made use of the support services provided by the repository staff or, on the contrary, the self-archiving route has been chosen for data deposit and publication. This latter consideration is linked to repository deposit policy that states mandatory and recommended metadata elements and qualifiers as well as quality check processes in place. In this sense, DIGITAL.CSIC metadata guidelines state a small number of mandatory metadata and permit researchers to complete a full deposit as long as such mandatory fields are covered. However, training and support services are made available for data creators to understand the importance of rich descriptions and implications as regards licenses and formats choice, for instance.

In this paper, by way of illustration, we are showcasing 2 completely different examples of datasets and how they score in FAIR EVA. The first one (10261/244749 dataset)\cite{10261_244749} is authored by a group of CSIC researchers that are participating in the Heritage Sciences ERIHs community \cite{ERIHs_web} within EOSC infrastructure. In this effort, DIGITAL.CSIC administrators have been brought into the conversation to agree on a descriptive model template that at the same time meets the repository policies and the needs and goals of the research community. On the other extreme of the ladder, we may find poorly described objects that pass the thresholds per indicators with rather low scores, most notably those referred to findability and interoperability. This is the case of the Biodiversity dataset entitled "Informed recruitment or the importance of taking stock" (10261/172425 dataset)\cite{10261_172425} which limits itself to the minimum metadata requested by DIGITAL.CSIC to finalize a submission, namely, author, title, date of publication, output type, data identifier, access status and language: however basic metadata are provided to support access, use and citation it is unlikely that the data are meaningful enough for users unless additional context is provided via author contact.

\begin{figure}[h]
\centering
\includegraphics[keepaspectratio=true,scale=0.2]{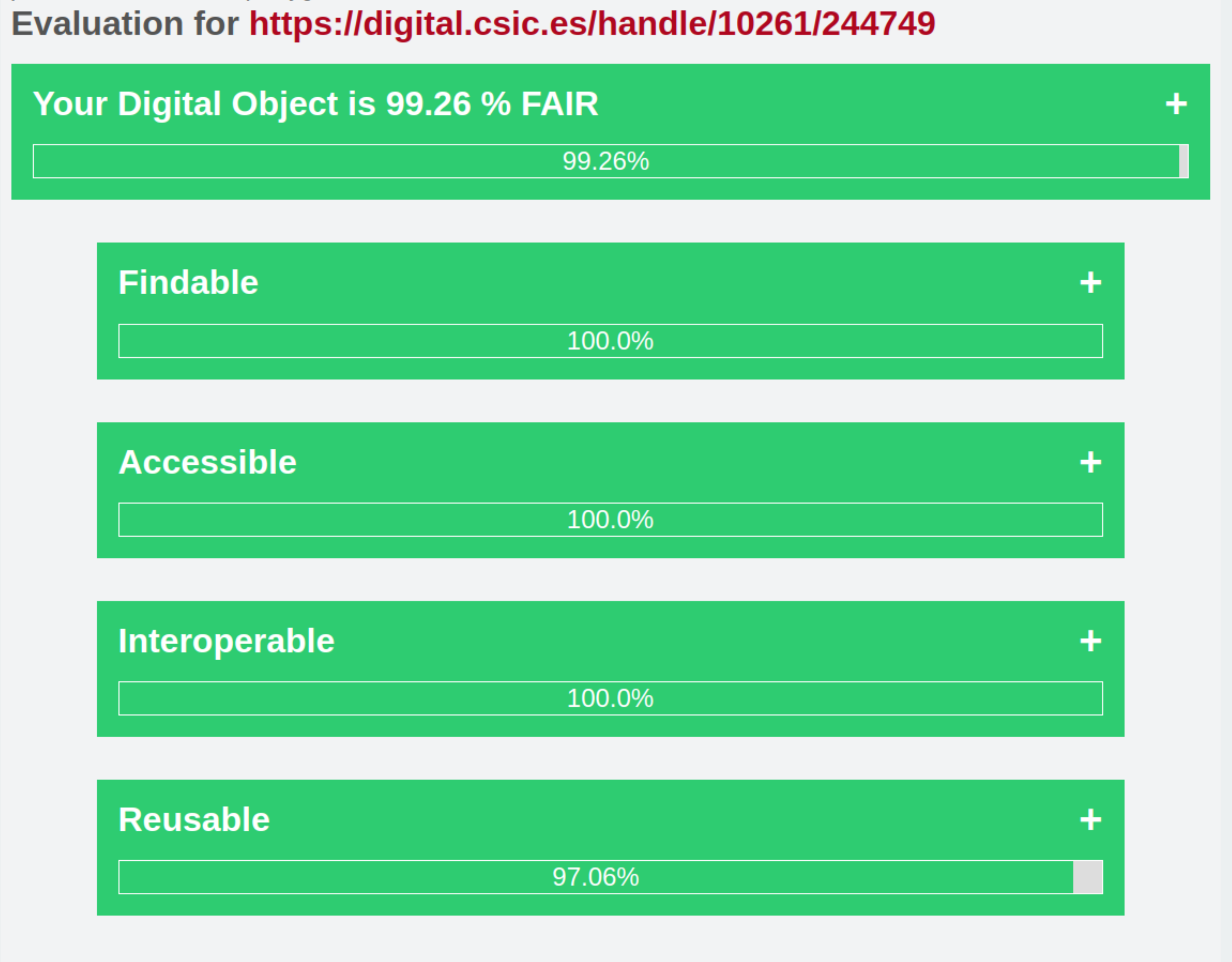}
\caption{Evaluation of object 10261/244749}
\label{fig:eval_244749}
\end{figure}

The results of the first dataset (10261/244749) are shown in Figure \ref{fig:eval_244749}. Taking into account the described Equation \ref{for:score}, this item has obtained a score of 99.26 \%, this is a good example of how a dataset can address the FAIR Principles. It includes a set of good practices of data management from the data producer point of view, including a unique and universal persistent identifier for the data, rich enough as well as qualified metadata, controlled vocabularies from the discipline involved (Getty Art and Architecture and UNESCO Vocabularies), relationships between the digital object and other contextual items or information, the presence of several types of unique identifiers (e.g. Wikidata) connected to searchable related resources etcetera. This set of characteristics are evaluated and validated with the test implementing the FAIR indicators in FAIR EVA, and it returns good results not only at generic or repository-dependent indicators but also for those relying on data producer practices.

\begin{figure}[h]
\centering
\includegraphics[keepaspectratio=true,scale=0.2]{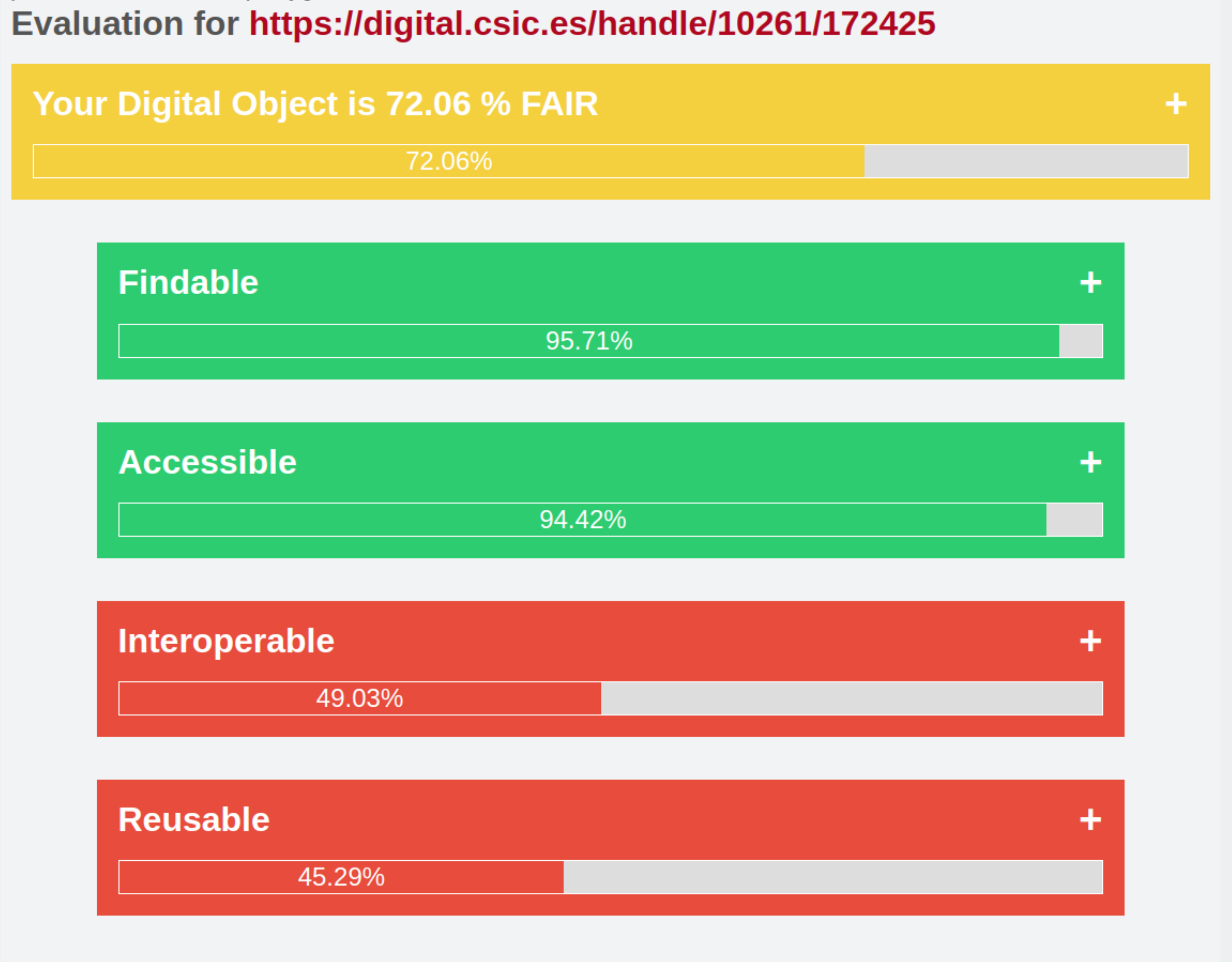}
\caption{Evaluation of object 10261/172425}
\label{fig:eval_172425}
\end{figure}

The results of the second dataset (10261/172425) are shown in Figure \ref{fig:eval_172425}. Taking into account the described Equation \ref{for:score}, the score of 72.06\% seems to indicate that the digital object is not a bad example. However, it relies mostly on the indicators derived from the repository infrastructure, which is supporting most of the FAIR characteristics by default. The dataset does not include any reference to other resources, nor use any controlled vocabulary to describe univocally some characteristics and it does not include rich enough qualified metadata, either. Although the difference is significant between both datasets, it does not fairly reflect the real gap in data quality and maturity. This is derived from the concept of the RDA indicators, which assigns more importance to some characteristics dependent on the repository that hosts the data.

\begin{figure}[h]
\centering
\includegraphics[keepaspectratio=true,scale=0.2]{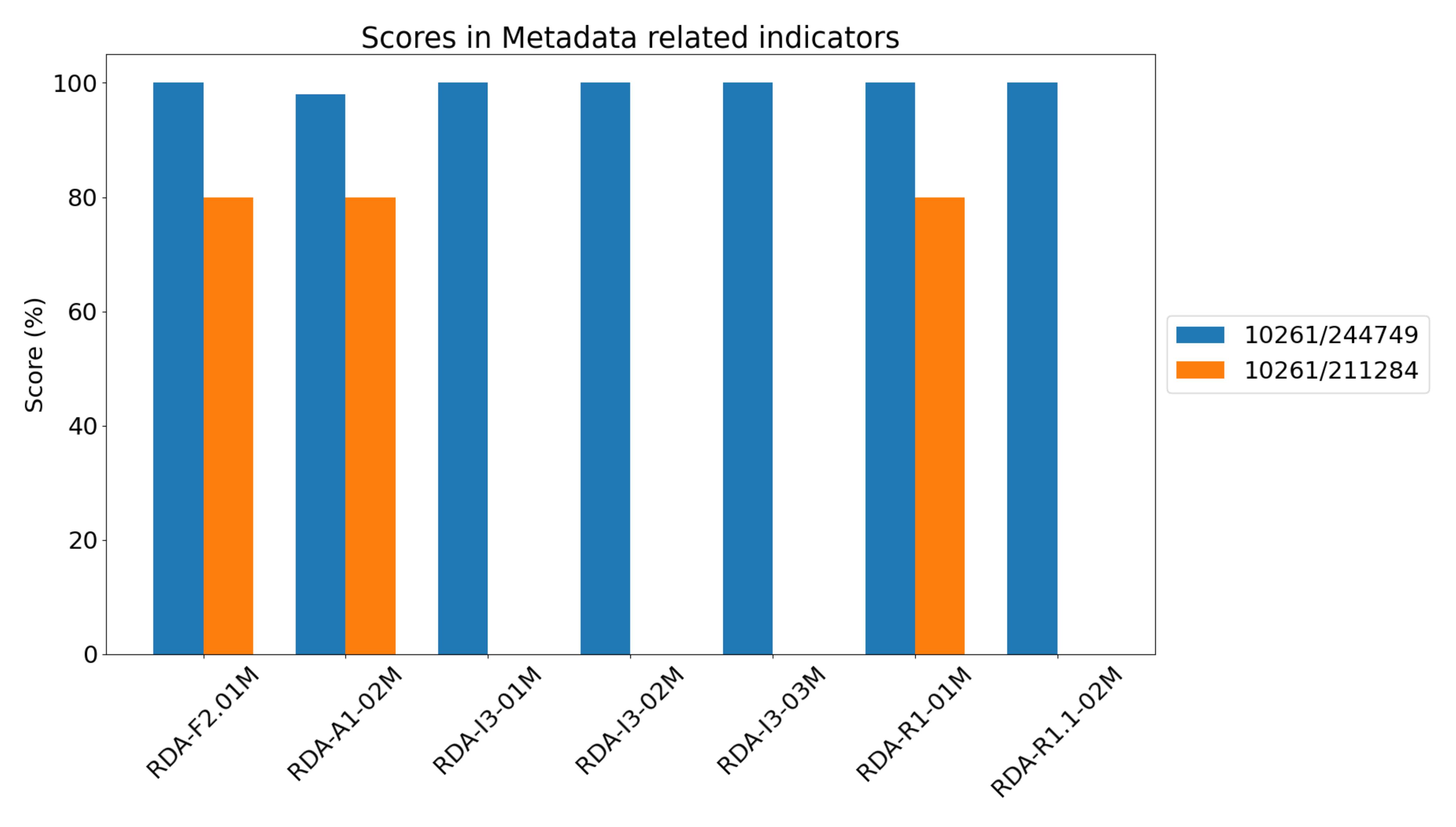}
\caption{Metrics of metadata-related indicators}
\label{fig:metadata_related}
\end{figure}

\begin{figure}[h]
\centering
\includegraphics[keepaspectratio=true,scale=0.2]{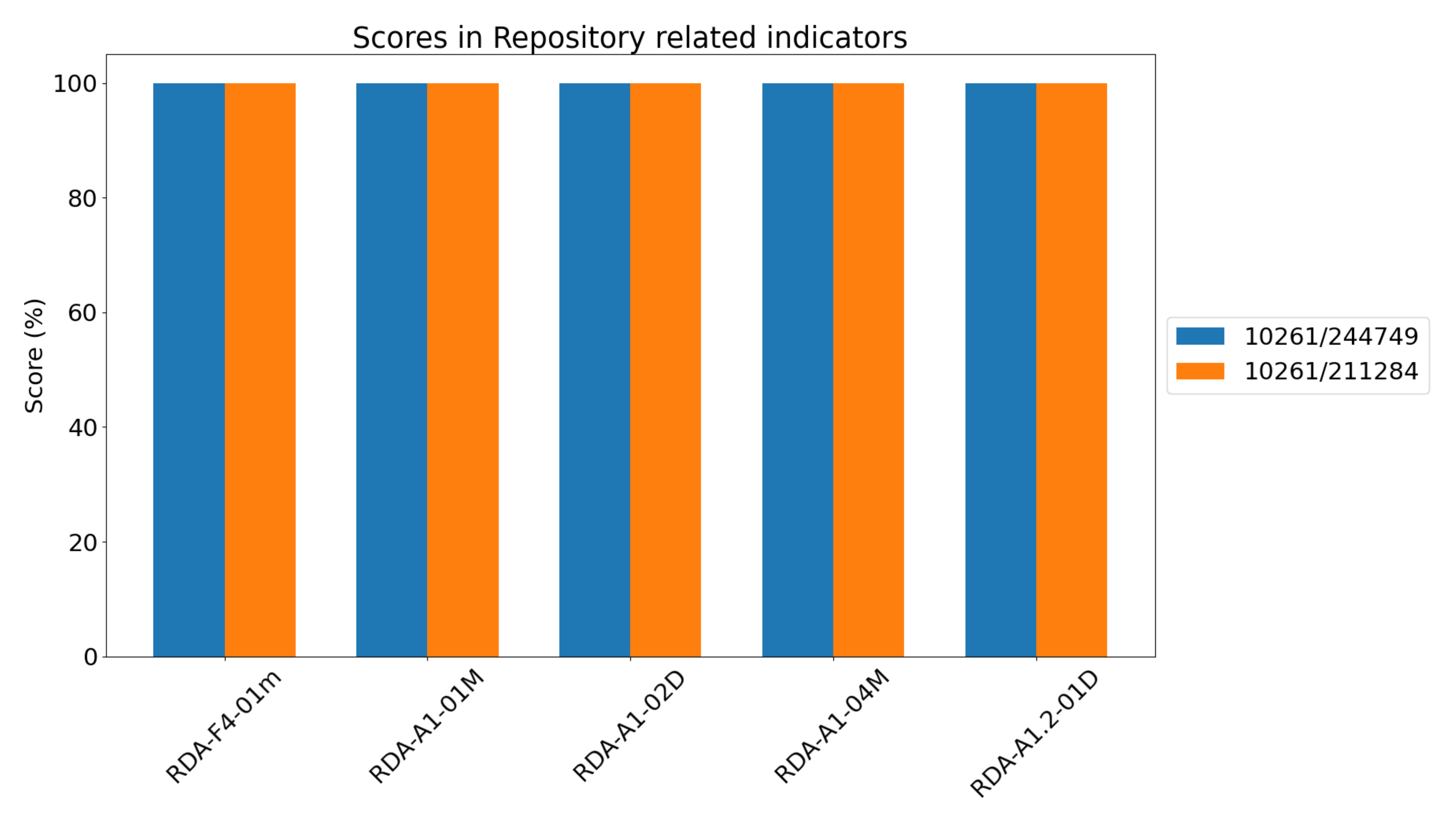}
\caption{Metrics of repository-related indicators}
\label{fig:repo_related}
\end{figure}

Analyzing in detail the results from the two tests showcased, an important characteristic for the RDA FAIR indicators and their technical implementation is revealed. Indeed, while some of the indicators largely depend on the data producer to be passed, some others trust repository features to be fulfilled. On one hand, Figure \ref{fig:repo_related} exposes a set of indicators that are based on how the repository works and which features are included in its technical infrastructure. This differentiation can help FAIR EVA users focus on those aspects that are within their reach for improvement, thus avoiding undue penalizations and unmeaningful feedback for users (see Figure \ref{fig:metadata_related}).

\section*{Discussion}

Although providing different weights to the level of importance of every single indicator (Essential, Important, Useful) is a good approach to emphasize some of the tests, some of them can be found repetitive, scoring more points for similar or almost equal checks. Also, as Table \ref{table:rda_levels} shows, the distribution of Essential, Important and Useful indicators is somehow unbalanced, with the greatest share of Essential indicators under F and A (7 and 8, respectively), while one finds 7 such indicators under I and just 5 under R. This distribution has a clear influence on the assessment outcomes, and in a way one may argue that the ultimate goal of FAIR (that of maximize the reuse of data) gets compromised. For example, within the list of Findable indicators, five of them (RDA-F1.01M, RDA-F1.01D, RDA-F1.02M, RDA-F1.02D and RDA-F3.01M) are tagged as "Essential" and are related to the presence of Persistent Identifiers, so it impacts significantly on the final score in an oversized way . Furthermore, most of the "Essential" indicators are defined in F and A, which are those more closely related to the repository infrastructure (see Figure \ref{fig:repo_related}). Therefore, depending on the data service, these tests will pass even if the user does not provide especially rich metadata or proper data, which also impact directly on the final score with significant weight.

\begin{table}[tbp]
        \centering
        \small
        \setlength\tabcolsep{2pt}
        \begin{tabular}{|c|c|c|c|}
\toprule 
 & \textbf{Essential (x2)} & \textbf{Important (x1.5)} & \textbf{Useful (x1)} \\
 \hline
\textbf{F indicators} & 7 & 0 & 0 \\
\hline
\textbf{A indicators} & 8 & 3 & 1 \\
\hline
\textbf{I indicators} & 0 & 7 & 5 \\
\hline
\textbf{R indicators} & 5 & 4 & 1 \\
\bottomrule
\end{tabular}
\caption{Distribution of priority types of RDA FAIR Indicators}
\label{table:rda_levels}
\end{table}

During the development of FAIR EVA and working in a rapidly changing context related to the evolution of the implementation and compliance assessment of FAIR Principles, different challenges have been detected and some problems solved. Research communities and institutions manage their data in very diverse environments and contexts, and the adoption of the FAIR Principles is not always easy and fluent. In this regard, some reflections on the practical implementation of FAIR Principles and RDA recommendations can be shared.

Multidisciplinary data repositories do present specific characteristics derived from their nature, mission, operational framework and wide spectrum of users. Indeed, it is important to understand and take the characteristics of the different data services into account. Institutional repositories community backs in a tradition of common guidelines and good practices that have favoured broad and early adoption of standards in digital object management, most notably by supporting features and services that enable discoverability, findability, access and interoperability. Mechanisms for adapting the FAIR assessment tools need to be enabled to adapt to different scenarios. In this context, the FAIR EVA tool can be adapted in such a way, by taking into account before running FAIR tests what metadata schema/s, controlled vocabularies/ontologies, default and optional properties and repository services/policies are made available to the user in a specific data infrastructure. Otherwise, it is of little use to both data producers and repository administrators to make use of FAIR assessment tools whose technical implementation is designed a priori to  assess against certain metadata schemas, ontologies, protocols and community standards only, thus excluding other valid and widely used standards by the community alongside other valid ways to check compliance and access data (for instance, the protocol being used to access (meta)data automatically).

FAIR EVA can be adapted to the objectives set by a given repository or data service as the goal of the tool is to make useful assessments that open the door to practical improvements.  For example, in the case of DIGITAL.CSIC, based on the RDA proposal, the "Indicator Level" can be Essential (x2), Important (x1.5) or Useful (x1). This is maintained both in the score obtained in each of the FAIR indicator groups and in the final score. To respect the customisation and adaptation feature, each plug-in can inherit or re-implement each test and a technical description is included in "Technical Implementation", while "Technical feedback" includes details of the test execution, incorporating information on the criteria set by each plug-in (e.g. metadata terms checked). Finally, one of the most important objectives of FAIR EVA is the ability to advise. The "Tips" section of the output includes information on how to improve each indicator, including links or references to documentation and learning materials.

Although the FAIR Principles are usually focused on data producers, research funding and research performing organizations, other stakeholders and actors deserve to play a more prominent role. First and foremost, repository software developers need to be brought into the conversation: first, open access repositories date back to the early 2000s (DSpace@MIT repository dates back to 2002) and the technological agenda of repository open source software tend to be community-driven. For several years, most focus and attention in the Open Access movement was heavily placed on publications and as a result, technological roadmaps were designed accordingly. However, there has been growing calls for next-generation repositories from the repository community itself for a while now (COAR Next generation working group). These new features include emerging protocols and technologies such as Signposting, ResourceSync, linked data integration, enhanced API REST capabilities, etcetera. They have only been made available in most recent software versions, which are still not broadly implemented within the community for many reasons, including intensive time and labour consuming and technically challenging migration processes as well as often lack of sufficient resources.

Complete and quicker integration of the latest developments in repository software through modular, scalable and flexible architectures will facilitate the broad adoption of FAIR Principles and their automatic assessment. Along these lines, it is noteworthy mentioning the outcomes of DSpace Visioning Group published at the end of 2021 \cite{dspaceProductVisioning} as clearly emphasize that “Future DSpace developments should be designed in ways that allow for easy integration with new and future tools or services in those areas most relevant to the community. In this respect, the development direction that DSpace is taking with its new API, together with incorporating features that support next-generation repository behaviours, can be a big step forward in supporting a wider range of integrations” and explicitly envisages “a future where DSpace will become certification-ready,” with “validation-ready” software solution for endorsed initiatives (e.g. certificates such as the CoreTrustSeal, metadata standards such as OpenAIRE and DataCite, or guiding principles such as the FAIR Data Principles and the TRUST Principles for digital repositories). This said, the agendas and work in progress of broadly used repository software need to be somehow represented in FAIR assessment tools to contribute to the global efforts for a FAIR ecosystem meaningfully.

If scientific communities want FAIR to become a full reality, it is clear that different scopes, contexts, services and research disciplines need to be taken into account. Starting from a set of generic indicators like the RDA ones, some interpretations and adaptations need to be done. FAIR EVA and its plug-in-based architecture allow the mechanisms to work with any data service with the minimum accessibility features.

\subsection*{Next steps for FAIR EVA at DIGITAL.CSIC}

% New
\subsubsection*{Semantics for transparency}
\begin{figure}[h]
\centering
\includegraphics[keepaspectratio=true,scale=0.3]{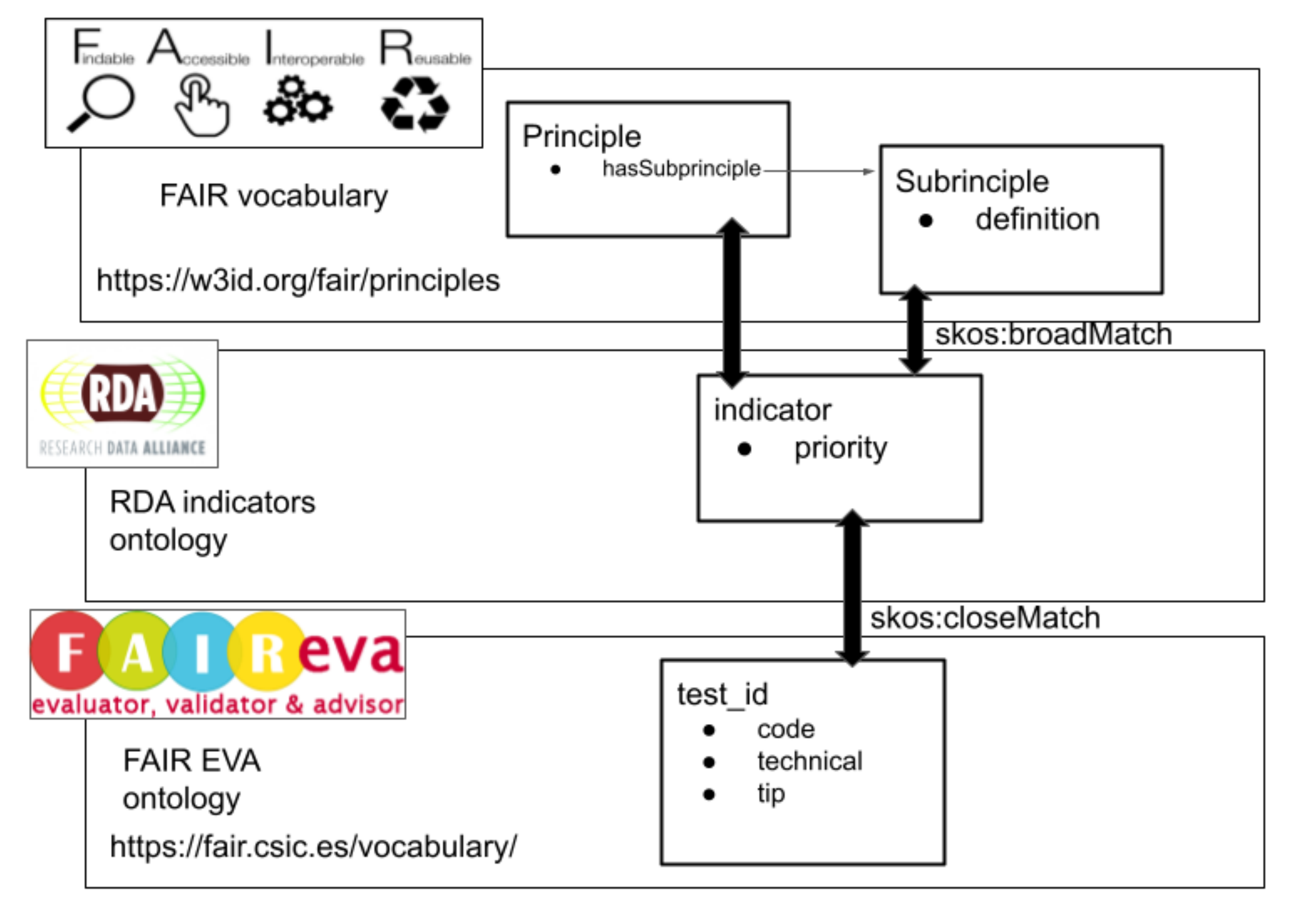}
\caption{Knowledge representation of test implementation and relationships with indicators and principles}
\label{fig:semantics}
\end{figure}

Living in a constantly changing FAIR environment, the assessment tools need to have a considerable degree of flexibility and adaptability, as well as enabling mechanisms to allow comparing. Apart from using same and compatible mechanisms to gather data and metadata as EOSC FAIR metrics Task Force suggests, FAIR EVA proposes the definition of the tests implementation and its relationships among metrics, indicators and the FAIR principles itself using semantic technology. Combining the use of already-existing ontologies defining FAIR Principles\cite{FAIR_vocabulary} and creating an ontology to define the test implementation, relationships can be detailed using the SKOS Simple Knowledge Organization System RDF schema, which indicates different types of relationships among knowledge representations systems. The Figure \ref{fig:semantics}shows the types of relationships to link the implementation of a specific test, the indicator that it implements and the FAIR Principle or sub-principle that it refers to. The ontology is available on the code repository.

Since RDA indicators are not published as semantic data officially yet, a temporary semantic description has been included in the FAIR EVA repository. It combines the description of the implemented metrics and their relationship with the indicators, which are also linked with the available FAIR vocabulary. Knowing the type of relationships among metrics implementation, indicators and the FAIR Principles themselves is essential to determine if different tools' outputs are not comparable.

Given the previous analysis, some potential improvements have been identified and some developments can be adapted both from the repository and the FAIR EVA sides:

\begin{itemize}
\item Enhance technical implementation of indicators more closely related to interoperability and reusability. In particular, by checking the existence of recommended practices to allow for replication/reproducibility of datasets as well as including new controlled vocabularies and ontologies into the scoring system as DIGITAL.CSIC enriches its semantic capabilities.
\item Factor preservation and web accessibility considerations around metadata and files.
\item Add new useful resources and tailor feedback for data producers/depositors depending on their discipline profile as well as additional recommendations for repository administrators. A first step in this direction, in order to provide data creators practical support around metadata guidelines on DIGITAL.CSIC, is the dedicated web page released last Fall\cite{fair_eva_support}.
\end{itemize}

We are also monitoring ongoing efforts led by several metadata related communities working to enhance interoperability amongst existing metadata schemas and advance new developments (for instance, by facilitating direct machine access to data files) to better align with FAIR Principles. In this regard, it is noteworthy to mention RDA Research Metadata Schemas Working Group that has collected and aligned crosswalks from 15 source research metadata schemas to Schema.org that can serve as a reference for data repositories when they develop their metadata crosswalks \cite{wu_mingfang_2022_6341481} as well as DDI-CDI metadata specification which is a model-driven, domain- and technology-neutral solution designed to facilitate the combination of data from diverse sources and across disciplines and possible use cases within EOSC framework \cite{gregory_arofan_2021_4707263}.

\section*{Data availability}
No new data was generated in this work.

\section*{Code availability}

The FAIR EVA source code developed in Python is fully open access (https://github.com/EOSC-synergy/FAIR\_eva), a running instance can be found in fair.csic.es and in full detail at DIGITAL.CSIC \cite{10261264113} [https://doi.org/10.20350/digitalCSIC/14559].
\bibliography{sample}

\begin{thebibliography}{10}
\urlstyle{rm}
\expandafter\ifx\csname url\endcsname\relax
  \def\url#1{\texttt{#1}}\fi
\expandafter\ifx\csname urlprefix\endcsname\relax\def\urlprefix{URL }\fi
\expandafter\ifx\csname doiprefix\endcsname\relax\def\doiprefix{DOI: }\fi
\providecommand{\bibinfo}[2]{#2}
\providecommand{\eprint}[2][]{\url{#2}}

\bibitem{Wilkinson2016}
\bibinfo{author}{Wilkinson, M.~D.} \emph{et~al.}
\newblock \bibinfo{journal}{\bibinfo{title}{{The FAIR Guiding Principles for
  scientific data management and stewardship}}}.
\newblock {\emph{\JournalTitle{Scientific Data}}} \textbf{\bibinfo{volume}{3}},
  \bibinfo{pages}{160018}, \url{10.1038/sdata.2016.18} (\bibinfo{year}{2016}).

\bibitem{eoscsecretariat_fair}
\bibinfo{title}{{FAIR Working Group | EOSCSecretariat}}.
\newblock
  \bibinfo{howpublished}{\url{https://www.eoscsecretariat.eu/working-groups/fair-working-group}}
  (\bibinfo{year}{2022}).

\bibitem{fair_taskforce}
\bibinfo{title}{{FAIR metrics and Data Quality | EOSC Association}}.
\newblock
  \bibinfo{howpublished}{\url{https://www.eosc.eu/advisory-groups/fair-metrics-and-data-quality}}
  (\bibinfo{year}{2022}).

\bibitem{doi/10.2777/54599}
\bibinfo{author}{Commission, E.}, \bibinfo{author}{for Research, D.-G.} \&
  \bibinfo{author}{Innovation}.
\newblock \emph{\bibinfo{title}{Turning FAIR into reality: final report and
  action plan from the European Commission expert group on FAIR data}}
  (\bibinfo{publisher}{Publications Office}, \bibinfo{year}{2018}).

\bibitem{Wilkinson2018}
\bibinfo{author}{Wilkinson, M.~D.} \emph{et~al.}
\newblock \bibinfo{journal}{\bibinfo{title}{{A design framework and exemplar
  metrics for FAIRness}}}.
\newblock {\emph{\JournalTitle{Scientific Data 2018 5:1}}}
  \textbf{\bibinfo{volume}{5}}, \bibinfo{pages}{1--4},
  \url{10.1038/sdata.2018.118} (\bibinfo{year}{2018}).

\bibitem{rda2020fair}
\bibinfo{author}{Group, R. F. D. M. M.~W.} \emph{et~al.}
\newblock \bibinfo{journal}{\bibinfo{title}{Fair data maturity model:
  specification and guidelines}}.
\newblock {\emph{\JournalTitle{Research Data Alliance. DOI}}}
  \textbf{\bibinfo{volume}{10}}, \url{10.15497/rda00050}
  (\bibinfo{year}{2020}).

\bibitem{fair_assist}
\bibinfo{title}{{FAIRassist.org | Help you discover resources to measure and
  improve FAIRness}}.
\newblock \bibinfo{howpublished}{\url{https://fairassist.org/}}
  (\bibinfo{year}{2022}).

\bibitem{bahim_christophe_2019_3629618}
\bibinfo{author}{Bahim, C.}, \bibinfo{author}{Dekkers, M.} \&
  \bibinfo{author}{Wyns, B.}
\newblock \bibinfo{title}{{Results of an Analysis of Existing FAIR Assessment
  Tools}}, \url{10.15497/rda00035} (\bibinfo{year}{2019}).

\bibitem{Knaisl}
\bibinfo{author}{Knaisl, V.} \& \bibinfo{author}{Such{\'{a}}nek, M.}
\newblock \bibinfo{title}{{FIP Wizard}}.
\newblock
  \bibinfo{howpublished}{\url{https://fip-wizard.readthedocs.io/en/latest/}}
  (\bibinfo{year}{2022}).

\bibitem{coarweb}
\bibinfo{title}{{COAR Community Framework for Good Practices in Repositories
  – COAR}}.
\newblock
  \bibinfo{howpublished}{\url{https://www.coar-repositories.org/coar-community-framework-for-good-practices-in-repositories/}}
  (\bibinfo{year}{2022}).

\bibitem{openaire_guide}
\bibinfo{title}{{OpenAIRE Guidelines — OpenAIRE Guidelines documentation}}.
\newblock
  \bibinfo{howpublished}{\url{https://guidelines.openaire.eu/en/latest/index.html}}
  (\bibinfo{year}{2022}).

\bibitem{base_guide}
\bibinfo{title}{{BASE - Bielefeld Academic Search Engine | Golden Rules for
  Repository Managers}}.
\newblock
  \bibinfo{howpublished}{\url{https://www.base-search.net/about/en/faq_oai.php}}
  (\bibinfo{year}{2022}).

\bibitem{doi/10.2777/986252}
\bibinfo{author}{Commission, E.}, \bibinfo{author}{for Research, D.-G.} \&
  \bibinfo{author}{Innovation}.
\newblock \emph{\bibinfo{title}{Six Recommendations for implementation of FAIR
  practice by the FAIR in practice task force of the European open science
  cloud FAIR working group}} (\bibinfo{publisher}{Publications Office},
  \bibinfo{year}{2020}).

\bibitem{Clarke2019}
\bibinfo{author}{Clarke, D.~J.} \emph{et~al.}
\newblock \bibinfo{journal}{\bibinfo{title}{Fairshake: Toolkit to evaluate the
  fairness of research digital resources}}.
\newblock {\emph{\JournalTitle{Cell Systems}}} \textbf{\bibinfo{volume}{9}},
  \bibinfo{pages}{417--421}, \url{10.1016/j.cels.2019.09.011}
  (\bibinfo{year}{2019}).

\bibitem{Bonello2022}
\bibinfo{author}{Bonello, J.}, \bibinfo{author}{Cachia, E.} \&
  \bibinfo{author}{Alfino, N.}
\newblock \bibinfo{journal}{\bibinfo{title}{Autofair-a portal for automating
  fair assessments for bioinformatics resources}}.
\newblock {\emph{\JournalTitle{Biochimica et Biophysica Acta (BBA) - Gene
  Regulatory Mechanisms}}} \textbf{\bibinfo{volume}{1865}},
  \bibinfo{pages}{194767}, \url{10.1016/J.BBAGRM.2021.194767}
  (\bibinfo{year}{2022}).

\bibitem{Wilkinson2019}
\bibinfo{author}{Wilkinson, M.~D.} \emph{et~al.}
\newblock \bibinfo{journal}{\bibinfo{title}{{Evaluating FAIR maturity through a
  scalable, automated, community-governed framework}}}.
\newblock {\emph{\JournalTitle{Scientific Data 2019 6:1}}}
  \textbf{\bibinfo{volume}{6}}, \bibinfo{pages}{1--12},
  \url{10.1038/s41597-019-0184-5} (\bibinfo{year}{2019}).

\bibitem{Homolak2020}
\bibinfo{author}{Homolak, J.}, \bibinfo{author}{Kodvanj, I.} \&
  \bibinfo{author}{Virag, D.}
\newblock \bibinfo{journal}{\bibinfo{title}{{Preliminary analysis of COVID-19
  academic information patterns: a call for open science in the times of closed
  borders}}}.
\newblock {\emph{\JournalTitle{Scientometrics}}}
  \textbf{\bibinfo{volume}{124}}, \bibinfo{pages}{2687--2701},
  \url{10.1007/s11192-020-03587-2} (\bibinfo{year}{2020}).

\bibitem{csicweb}
\bibinfo{title}{{Consejo Superior de Investigaciones Científicas}}.
\newblock \bibinfo{howpublished}{\url{https://www.csic.es/en}}
  (\bibinfo{year}{2023}).

\bibitem{stall2019make}
\bibinfo{author}{Stall, S.} \emph{et~al.}
\newblock \bibinfo{title}{Make scientific data fair}.
\newblock
  \bibinfo{howpublished}{\url{https://www.nature.com/articles/d41586-019-01720-7}}
  (\bibinfo{year}{2019}).

\bibitem{10.18665/sr.316121}
\bibinfo{author}{Ruediger, D.} \emph{et~al.}
\newblock \bibinfo{journal}{\bibinfo{title}{{Big Data Infrastructure at the
  Crossroads: Support Needs and Challenges for Universities}}}.
\newblock {\emph{\JournalTitle{Ithaka S+R}}} \url{10.18665/SR.316121}
  (\bibinfo{year}{2021}).

\bibitem{copernicusData2019}
\bibinfo{title}{Copernicus sentinel data access 2019 annual report}.
\newblock
  \bibinfo{howpublished}{\url{https://scihub.copernicus.eu/twiki/pub/SciHubWebPortal/AnnualReport2019/COPE-SERCO-RP-20-0570_-_Sentinel_Data_Access_Annual_Report_Y2019_v1.0.pdf}}
  (\bibinfo{year}{2021}).

\bibitem{merkel2014docker}
\bibinfo{author}{Merkel, D.}
\newblock \bibinfo{journal}{\bibinfo{title}{Docker: lightweight linux
  containers for consistent development and deployment}}.
\newblock {\emph{\JournalTitle{Linux journal}}}
  \textbf{\bibinfo{volume}{2014}}, \bibinfo{pages}{2} (\bibinfo{year}{2014}).

\bibitem{Jenkins}
\bibinfo{title}{{Jenkins User Documentation}}.
\newblock \bibinfo{howpublished}{\url{https://www.jenkins.io/doc/}}
  (\bibinfo{year}{2022}).

\bibitem{OrvizFernandez2020}
\bibinfo{author}{{Orviz Fern{\'{a}}ndez}, P.} \emph{et~al.}
\newblock \bibinfo{journal}{\bibinfo{title}{{Software Quality Assurance in
  INDIGO-DataCloud Project: a Converging Evolution of Software Engineering
  Practices to Support European Research e-Infrastructures}}}.
\newblock {\emph{\JournalTitle{Journal of Grid Computing 2020 18:1}}}
  \textbf{\bibinfo{volume}{18}}, \bibinfo{pages}{81--98},
  \url{10.1007/S10723-020-09509-Z} (\bibinfo{year}{2020}).

\bibitem{SSHOPENCLOUD_web}
\bibinfo{title}{Sshopencloud | social sciences \& humanities open cloud}.
\newblock \bibinfo{howpublished}{\url{https://sshopencloud.eu/}}
  (\bibinfo{year}{2022}).

\bibitem{ELIXIR_web}
\bibinfo{title}{Elixir | a distributed infrastructure for life-science
  information}.
\newblock \bibinfo{howpublished}{\url{https://elixir-europe.org/}}
  (\bibinfo{year}{2022}).

\bibitem{ENVRI_web}
\bibinfo{title}{Envri community – studying the environment today to tackle
  the challenges of tomorrow}.
\newblock \bibinfo{howpublished}{\url{https://envri.eu/}}
  (\bibinfo{year}{2022}).

\bibitem{eosc_rules}
\bibinfo{title}{{EOSC rules of participation - Publications Office of the EU}}.
\newblock
  \bibinfo{howpublished}{\url{https://op.europa.eu/en/publication-detail/-/publication/a96d6233-554e-11eb-b59f-01aa75ed71a1/language-en/format-PDF/source-184432576}}
  (\bibinfo{year}{2022}).

\bibitem{DeSompel2015}
\bibinfo{author}{{De Sompel}, H.~V.} \& \bibinfo{author}{Nelson, M.~L.}
\newblock \bibinfo{journal}{\bibinfo{title}{{Reminiscing about 15 years of
  interoperability efforts}}}.
\newblock {\emph{\JournalTitle{D-Lib Magazine}}} \textbf{\bibinfo{volume}{21}},
  \bibinfo{pages}{1--1}, \url{10.1045/NOVEMBER2015-VANDESOMPEL}
  (\bibinfo{year}{2015}).

\bibitem{Lin2020}
\bibinfo{author}{Lin, D.} \emph{et~al.}
\newblock \bibinfo{title}{{The TRUST Principles for digital repositories}},
  \url{10.1038/s41597-020-0486-7} (\bibinfo{year}{2020}).

\bibitem{Hahnel2020}
\bibinfo{author}{Hahnel, M.} \& \bibinfo{author}{Valen, D.}
\newblock \bibinfo{journal}{\bibinfo{title}{{How to (Easily) Extend the
  FAIRness of Existing Repositories}}}.
\newblock {\emph{\JournalTitle{Data Intelligence}}}
  \textbf{\bibinfo{volume}{2}}, \bibinfo{pages}{192--198},
  \url{10.1162/dint_a_00041} (\bibinfo{year}{2020}).

\bibitem{Bahim2019}
\bibinfo{author}{Bahim, C.}, \bibinfo{author}{Dekkers, M.} \&
  \bibinfo{author}{Wyns, B.}
\newblock \bibinfo{journal}{\bibinfo{title}{{Results of an Analysis of Existing
  FAIR Assessment Tools}}}.
\newblock {\emph{\JournalTitle{Zenodo}}} \url{10.15497/RDA00035}
  (\bibinfo{year}{2019}).

\bibitem{Sun2022}
\bibinfo{author}{Sun, C.}
\newblock \bibinfo{journal}{\bibinfo{title}{{A comprehensive comparison of
  automated FAIRness Evaluation Tools}}}.
\newblock {\emph{\JournalTitle{CEUR WS}}} \url{10.1594/PANGAEA.908011}
  (\bibinfo{year}{2022}).

\bibitem{synergyd35}
\bibinfo{author}{Gómez, F.~A.}, \bibinfo{author}{Gomes, J.},
  \bibinfo{author}{Bernal, I.}, \bibinfo{author}{Steinhoff, W.} \&
  \bibinfo{author}{Tykhonov, V.}
\newblock \bibinfo{journal}{\bibinfo{title}{Eosc-synergy eu deliverable d3.5:
  Final report on technical framework for eosc fair data principles
  implementation}}.
\newblock {\emph{\JournalTitle{EOSC-Synergy Deliverables}}}
  \url{10.20350/digitalCSIC/14888} (\bibinfo{year}{2022}).

\bibitem{fairChecker}
\bibinfo{author}{Rosnet, T.}, \bibinfo{author}{Gaignard, A.} \&
  \bibinfo{author}{Devignes, M.-D.}
\newblock \bibinfo{title}{Fair-checker}.
\newblock
  \bibinfo{howpublished}{\url{https://fair-checker.france-bioinformatique.fr/}}
  (\bibinfo{year}{2023}).

\bibitem{Devaraju2021}
\bibinfo{author}{Devaraju, A.} \& \bibinfo{author}{Huber, R.}
\newblock \bibinfo{journal}{\bibinfo{title}{An automated solution for measuring
  the progress toward fair research data}}.
\newblock {\emph{\JournalTitle{Patterns}}} \textbf{\bibinfo{volume}{2}},
  \bibinfo{pages}{100370}, \url{10.1016/j.patter.2021.100370}
  (\bibinfo{year}{2021}).

\bibitem{synergy_web}
\bibinfo{title}{{EOSC synergy – Building capacity, developing capability}}.
\newblock \bibinfo{howpublished}{\url{https://www.eosc-synergy.eu/}}
  (\bibinfo{year}{2022}).

\bibitem{re3data}
\bibinfo{title}{{Home | re3data.org}}.
\newblock \bibinfo{howpublished}{\url{https://www.re3data.org/}}
  (\bibinfo{year}{2022}).

\bibitem{Mandato}
\bibinfo{title}{{Mandato Institucional de Acceso Abierto | DIGITAL.CSIC}}.
\newblock
  \bibinfo{howpublished}{\url{https://digital.csic.es/handle/10261/179077}}
  (\bibinfo{year}{2022}).

\bibitem{mandatoMonitor}
\bibinfo{author}{CSIC}.
\newblock \bibinfo{title}{Digital.csic: monitor del mandato de acceso abierto
  del csic - home}.
\newblock
  \bibinfo{howpublished}{\url{https://digital.csic.es/sites/monitor_mandato_oa_csic/index.html}}
  (\bibinfo{year}{2023}).

\bibitem{FAIR_eva_tech}
\bibinfo{author}{Aguilar, F.} \& \bibinfo{author}{Bernal, I.}
\newblock \bibinfo{title}{Fair\_eva/technical\_implementation.md at main ·
  eosc-synergy/fair\_eva}.
\newblock
  \bibinfo{howpublished}{\url{https://github.com/EOSC-synergy/FAIR_eva/blob/main/docs/technical_implementation.md}}
  (\bibinfo{year}{2022}).

\bibitem{flask}
\bibinfo{author}{Grinberg, M.}
\newblock \emph{\bibinfo{title}{Flask web development: developing web
  applications with python}} (\bibinfo{publisher}{O'Reilly Media, Inc.},
  \bibinfo{year}{2018}).

\bibitem{openApi}
\bibinfo{title}{{OpenAPI Specification v3.1.0 | Introduction, Definitions, \&
  More}}.
\newblock \bibinfo{howpublished}{\url{https://spec.openapis.org/oas/v3.1.0}}
  (\bibinfo{year}{2022}).

\bibitem{fairEvaRepo}
\bibinfo{author}{Aguilar, F.}
\newblock \bibinfo{title}{Eosc-synergy/fair\_eva}.
\newblock
  \bibinfo{howpublished}{\url{hhttps://github.com/EOSC-synergy/FAIR\_eva}}
  (\bibinfo{year}{2019}).

\bibitem{ayuda_fair_eva}
\bibinfo{author}{CSIC}.
\newblock \bibinfo{title}{Ayuda fair eva - home}.
\newblock
  \bibinfo{howpublished}{\url{https://digital.csic.es/sites/ayuda\_fair\_data/index.html}}
  (\bibinfo{year}{2022}).

\bibitem{10261_244749}
\bibinfo{author}{Oujja, M.} \emph{et~al.}
\newblock \bibinfo{title}{Dataset for the paper "multiphoton excitation
  fluorescence microscopy and spectroscopic multianalytical approach for
  characterization of historical glass grisailles. talanta 230, 122314"},
  \url{10.20350/digitalCSIC/13919} (\bibinfo{year}{2021}).

\bibitem{ERIHs_web}
\bibinfo{title}{European research infrastructure for heritage science}.
\newblock \bibinfo{howpublished}{\url{https://www.e-rihs.eu/}}
  (\bibinfo{year}{2023}).

\bibitem{10261_172425}
\bibinfo{author}{Genovart, M.} \& \bibinfo{author}{Oro, D.}
\newblock \bibinfo{title}{Informed recruitment or the importance of taking
  stock}.
\newblock
  \bibinfo{howpublished}{\url{https://digital.csic.es/handle/10261/172425}},
  \url{10.20350/DIGITALCSIC/8581} (\bibinfo{year}{2018}).

\bibitem{dspaceProductVisioning}
\bibinfo{title}{Dspace product visioning group - dspace - lyrasis wiki}.
\newblock
  \bibinfo{howpublished}{\url{https://wiki.lyrasis.org/display/DSPACE/DSpace+Product+Visioning+Group?preview=/199525834/230817992/dspace-product-visioning-working-group-report-20211206.pdf}}
  (\bibinfo{year}{2022}).

\bibitem{FAIR_vocabulary}
\bibinfo{author}{Kuhn, T.} \& \bibinfo{author}{Dumontier, M.}
\newblock \bibinfo{title}{Fair vocabulary}.
\newblock
  \bibinfo{howpublished}{\url{https://w3id.org/fair/principles/terms/A}}
  (\bibinfo{year}{2022}).

\bibitem{fair_eva_support}
\bibinfo{title}{{Ayuda FAIR EVA}}.
\newblock
  \bibinfo{howpublished}{\url{https://digital.csic.es/sites/ayuda_fair_data/index.html}}
  (\bibinfo{year}{2022}).

\bibitem{wu_mingfang_2022_6341481}
\bibinfo{author}{Wu, M.} \emph{et~al.}
\newblock \bibinfo{title}{{A Collection of Crosswalks from Fifteen Research
  Data Schemas to Schema.org}}, \url{10.15497/RDA00069} (\bibinfo{year}{2022}).

\bibitem{gregory_arofan_2021_4707263}
\bibinfo{author}{Gregory, A.}, \bibinfo{author}{Hodson, S.} \&
  \bibinfo{author}{Wackerow, J.}
\newblock \bibinfo{title}{{The Role of DDI-CDI in EOSC: Possible Uses and
  Applications}}, \url{10.5281/zenodo.4707263} (\bibinfo{year}{2021}).

\bibitem{10261264113}
\bibinfo{author}{Aguilar-Gómez, F.}
\newblock \bibinfo{title}{Fair eva (evaluator, validator \& advisor)},
  \url{https://doi.org/10.20350/digitalCSIC/14559} (\bibinfo{year}{2022}).

\end{thebibliography}

\section*{Acknowledgements}

The project leading to this application, EOSC-Synergy, has received funding from the European Union's Horizon 2020 research and innovation programme under grant agreement No 857647.

\subsection*{Author information}
These authors contributed equally: Fernando Aguilar Gómez, Isabel Bernal.

\subsubsection*{Affiliations}
\textbf{Spanish National Research Council (CSIC), Instituto de Física de Cantabria (IFCA), Santander, Spain}
Fernando Aguilar Gómez

\textbf{Spanish National Research Council (CSIC), DIGITAL.CSIC, Unidad de Recursos de Información Científica para la Investigación, Madrid, Spain}
Isabel Bernal

\subsubsection*{Contributions}
Fernando Aguilar Gómez conceived the work. Isabel Bernal provided her expertise in academic repositories and data management best practices and elaborated on DIGITAL.CSIC infrastructure and services, and its technological features and metadata guidelines in order to enable this FAIR EVA implementation. Fernando Aguilar Gómez implemented the FAIR principles in FAIR EVA and he is the main developer. Both authors have contributed to the look and feel and structure of the FAIR Eva site (fair.csic.es) and thank Juan Román Molina of DIGITAL.CSIC Team for their support. 

\section*{Ethics declarations}
\subsection*{Competing interests}

The authors declare no competing interests.

\end{document}